\documentclass[aps,pra,amsmath,twocolumn,amssymb,superscriptaddress]{revtex4-1}

\usepackage{amssymb,amsfonts,amsmath}

\usepackage{color}
\usepackage[apple mac]{inputenc}
\usepackage[english]{babel}
\usepackage{times}
\usepackage{latexsym}
\usepackage{fancyhdr}
\usepackage{verbatim}
\usepackage{graphicx}
\usepackage{epsf}
\usepackage{subfigure}
\usepackage{bm}
\usepackage{epstopdf}\usepackage[colorlinks=true , citecolor=blue,urlcolor=blue]{hyperref}
\usepackage{units}

\usepackage[
top    = 1.5cm,
bottom = 1.50cm,
left   = 1.5cm,
right  = 0.9cm]{geometry}

\definecolor{Nathanblue}{rgb}{0.96,0.24,0.00}

\definecolor{Nathanbluetitle}{rgb}{0.,0.24,0.51}
\newcommand{\bluetitle}{\color{Nathanbluetitle}}

\definecolor{orange}{rgb}{0.96,0.24,0.00}

\def\be{\begin{equation}}
\def\ee{\end{equation}}
\def\bs#1{\boldsymbol{#1}}

\begin{document}

\title{{\bluetitle  Periodically-driven quantum matter: the case of resonant modulations}}

\author{N. Goldman}
\email[]{nathan.goldman@lkb.ens.fr}
\email[]{ngoldman@ulb.ac.be}
\affiliation{Laboratoire Kastler Brossel, Coll\`ege de France, CNRS, ENS-PSL Research University, UPMC-Sorbonne Universit\'es, 11 place Marcelin Berthelot, 75005, Paris, France}
\affiliation{CENOLI, Facult{\'e} des Sciences, Universit{\'e} Libre de Bruxelles (U.L.B.), B-1050 Brussels, Belgium}

\author{J. Dalibard}
\email[]{jean.dalibard@lkb.ens.fr}
\affiliation{Laboratoire Kastler Brossel, Coll\`ege de France, CNRS, ENS-PSL Research University, UPMC-Sorbonne Universit\'es, 11 place Marcelin Berthelot, 75005, Paris, France}

\author{M. Aidelsburger}
\email[]{Monika.Aidelsburger@physik.uni-muenchen.de}
\affiliation{Fakult\"at f\"ur Physik, Ludwig-Maximilians-Universit\"at, Schellingstrasse 4, 80799 M\"unchen, Germany}
\affiliation{Max-Planck-Institut f\"ur Quantenoptik, Hans-Kopfermann-Strasse 1, 85748 Garching, Germany}

\author{N. R. Cooper}
\email[]{nrc25@cam.ac.uk}
\affiliation{T.C.M. Group, Cavendish Laboratory, J.J. Thomson Avenue, Cambridge CB3 0HE, United Kingdom}

\date{\today}

\begin{abstract}
Quantum systems can  show qualitatively new forms of behavior when they are driven by fast time-periodic modulations. In the limit of large driving frequency, the long-time dynamics of such systems  can often be described by a time-independent effective Hamiltonian, which is generally identified through a perturbative treatment. Here, we present a general formalism that describes time-modulated physical systems, in which the driving frequency is large, but resonant with respect to energy spacings inherent to the system at rest. Such a situation is currently exploited in optical-lattice setups, where superlattice  (or Wannier-Stark-ladder) potentials are resonantly modulated so as to control the tunneling matrix elements between lattice sites, offering a powerful method to generate artificial fluxes for cold-atom systems. The formalism developed in this work identifies the basic ingredients needed to generate interesting flux patterns and band structures using resonant modulations. Additionally, our approach allows for a simple description of the micro-motion underlying the dynamics; we illustrate its characteristics based on diverse dynamic-lattice configurations. It is shown that the impact of the micro-motion  on physical observables strongly depends on the implemented scheme, suggesting that a theoretical description in terms of the effective Hamiltonian alone is generally not sufficient to capture the full time-evolution of the system.

\end{abstract}

\maketitle

\section{Introduction}

Subjecting a material to fast-oscillating fields constitutes a versatile scenario to reach and manipulate unusual quantum phases in solid-state laboratories, such as high-temperature superconductors \cite{Fausti:2011,Forst:2014,Matsunaga:2014} and topological quantum states of matter \cite{Cayssol:2013, Kitagawa:2010,Lindner:2011,Calvo:2011,SuarezMorell:2012,Tong:2012,Wang:2013,GomezLeon:2013,Delplace:2013,GomezLeon:2013,Rudner:2013,Grushin:2014,PerezPiskunow:2014,Wang:2014,Lindner:2013}. This approach is rooted in the fact that the dynamics associated with time-dependent Hamiltonians can be well captured by a time-\emph{independent} effective Hamiltonian $\hat H_{\text{eff}}$, in the limit of infinitely small driving period $T \!\ll\! t_{\text{ch}}$, where $t_{\text{ch}}$ denotes a typical time-scale for the dynamical properties under scrutiny, see Refs. \cite{Maricq:1982,Rahav:2003,Kitagawa:2010,Arimondo2012,Cayssol:2013,goldmandalibard,Bukov:2014_review}. In this picture, the energy spectrum of the static system is replaced by the  (Floquet) spectrum associated with $\hat H_{\text{eff}}$, which can potentially present interesting features, such as topological properties. 

The idea to enrich a physical system by designing a time-modulation protocol has inspired several other fields of research. It was recently applied to photonic crystals \cite{Rechtsman:2013}, ion traps \cite{Bermudez:2011prl,Bermudez:2012njp} and cold-atom setups \cite{Lignier:2007,Eckardt:2007epl,Lim2008,Eckardt2010,Hemmerich2010,Jiang:2011,Struck:2011,Aidelsburger:2011,Arimondo2012,Struck2012,Hauke:2012,Xu2013,Anderson2013,Struck:2013,Aidelsburger:2013,Aidelsburger:2013b,Ketterle:2013,Kennedy:2013,goldmandalibard,Baur:2014ux,Atala:2014uc,Reichl:2014,Jotzu:2014,Aidelsburger:2014,Zheng:2014,StruckSimonet:2014,Garcia:2014,Luo:2015}. In particular, optical-lattice potentials for cold atoms \cite{Bloch2008a} are ideally suited for generating a wide family of time-dependent potentials. These modulated potentials recently led to the experimental realization of effective magnetic fields in square \cite{Aidelsburger:2011,Aidelsburger:2013,Aidelsburger:2013b,Ketterle:2013,Aidelsburger:2014}, triangular \cite{Struck:2011,Struck:2013}  and honeycomb lattices \cite{Jotzu:2014}. Such arrangements already revealed striking phenomena, including  frustrated magnetism \cite{Struck:2011,Struck:2013}, chiral currents \cite{Atala:2014uc}, signatures of the Berry curvature \cite{Jotzu:2014} and the measurement of a non-trivial Chern number \cite{Aidelsburger:2014}. 

Time-modulated quantum systems can be classified into two distinct families. In the first class, the driving frequency $\omega \!=\! 2 \pi / T$ is arbitrarily large and it is \emph{off-resonant} with respect to any energy separation $\Delta$ intrinsic to the static system. The shaken optical lattices of Refs. \cite{Lignier:2007,Struck:2011,Struck2012,Struck:2013,Jotzu:2014}  belong to this first category. The second class concerns systems involving a \emph{resonant} modulation. The experimental setups of Refs. \cite{Aidelsburger:2011,Aidelsburger:2013,Aidelsburger:2013b,Ketterle:2013,Aidelsburger:2014} belong to this second class, where a superlattice (or a Wannier-Stark ladder) with energy offsets $\pm \Delta$ between neighboring sites was combined with a resonant modulation [with frequency $\omega \!=\! \Delta/\hbar$] to induce tunneling over the lattice in a controllable manner.

In Ref.~\cite{goldmandalibard}, a general formalism that analyzes periodically-driven quantum systems was developed, with a view to identifying realistic schemes leading to interesting band structures. This approach, which generalizes the work of Ref.~\cite{Rahav:2003}, provides a systematic method to obtain an unambiguous effective Hamiltonian $\hat H_{\text{eff}}$ ruling the long-time dynamics, together with a so-called ``kick" operator $\hat K (t)$ describing the micro-motion [i.e. the rapid motion undergone within one period of the driving].  Although general, the formalism presented in Ref.~\cite{goldmandalibard} was dedicated to time-dependent systems involving \emph{off-resonant} modulations. 

In this work, we extend the formalism of Ref.~\cite{goldmandalibard} so as to include the case of resonant modulations. We obtain general expressions for the effective Hamiltonian and kick operators, and apply them to diverse schemes involving two-dimensional superlattices or Wannier-Stark ladders. A particular emphasis is set on the effective-magnetic-flux configurations generated in modulated optical lattices, such as those leading to Chern bands with large flatness ratio. The latter band properties are particularly intriguing, as they could be exploited to produce fractional Chern insulators~\cite{Bergholtz:2013,Parameswaran:2013,Neupert:2014wt} with cold-atom setups, see also Refs.~\cite{Sorensen2005,Hafezi2007,Palmer:2008,Moller:2009,CooperDalibard:2012,Yao:2012fn,Yao:2013eg,Peter:2014wu,Sterdyniak:2014}. Analytical expressions are obtained to describe the micro-motion in these schemes; this allows one to predict the pattern and time-dependence of momentum distributions, as revealed by time-of-flight images.

The paper is structured as follows. In Section \ref{section_general}, we propose a general method to extend the formalism of Ref. \cite{goldmandalibard} to the case of resonant driving. We show how to handle this subtle situation,  where the static Hamiltonian now includes energy offset terms $\sim \!\Delta \!=\! \hbar \omega$, which diverge with the driving frequency $\omega\! \rightarrow \!\infty$.  Section \ref{section_magnus} discusses how additional (possibly time-dependent) diverging terms can be handled within this formalism; in particular,  Section \ref{section_magnus} analyzes the ``strong-driving" regime of time-modulated systems, where the modulation strength is of the order of the driving frequency. Section \ref{section_nigel}  applies the formalism to the case of general superlattices involving two sites per unit cell; it discusses the effective dynamics and micro-motion based on a momentum-space picture. We then study in Section \ref{section_one_harmonics} the case of two-dimensional square superlattices and Wannier-Stark ladders  with energy offsets $\pm \Delta$ between neighboring sites; in particular, we discuss the effective flux patterns that can be generated in such configurations. These results are illustrated in Section \ref{section_uniform}, which analyses specific experimental schemes. An emphasis is set on the different micro-motions associated with these schemes. The possibility to locally control the tunneling matrix elements and flux patterns is also presented. The Section \ref{section_more} extends the results of Section \ref{section_one_harmonics} to the more general case where the superlattice contains high-order offsets $\Delta \times \text{integer}$, which can potentially lead to even richer flux patterns when modulating the system with higher harmonics $\omega \times \text{integer}$. We conclude with final remarks and outlooks  in Section \ref{section_conclusions}.

\section{General framework}\label{section_general}

We consider the behavior of quantum systems subjected to time-periodic 
Hamiltonians of the form
\be
\hat H (t) = \sum_{j=- \infty}^{\infty} \hat H^{(j)} e^{i j \omega t} , \label{eq:h}
\ee
and we define $T \!=\! 2\pi/\omega$ as the period.  In Ref.~\onlinecite{goldmandalibard}, it was shown 
that the dynamical behavior at large frequency $\omega\!\to\!
\infty$ can be uniquely represented in terms of an effective
(time-independent) Hamiltonian $\hat H_{\text{eff}}$,  and a time-periodic ``kick'' operator $\hat K(t)$ with a zero time average over one period. In this picture, the time-evolution operator is defined and partitioned as
\begin{align}
\psi (t) \!=\! \hat U(t;t_0) \psi (t_0) ; \,\,  \hat U(t;t_0) \!=\!e^{-i\hat K(t)}e^{- \frac{i}{\hbar} (t - t_0)  \hat H_{\rm eff}}e^{i\hat K(t_0)},\label{partition}
\end{align} 
where the kick operator $\hat K(t)$ both describes the initial kick $\exp [i \hat K (t_0)]$ acting on the system at the initial time $t_0$, as well as the micro-motion $\exp [-i \hat K (t)]$ undergone within a period ($t \ne T \times \text{integer}$). The effective Hamiltonian $\hat H_{\text{eff}}$ and kick operator $\hat K(t)$ can be systematically computed using a perturbative expansion \cite{goldmandalibard,Rahav:2003} in powers of $1/\omega$, assuming that the Hamiltonian $\hat{H}(t)$ remains finite in the limit $\omega \to \infty$.

Here, we generalize this approach to allow the static component $\hat{H}^{(0)}$ to include
terms proportional to $\omega$, which therefore also diverge in the
limit $\omega\to \infty$. This situation occurs in lattice systems using {\it resonant} restoration of tunneling \cite{Eckardt:2007epl,Kolovsky:2011,Bermudez:2011prl}, a technique which has been recently implemented \cite{Aidelsburger:2011,Aidelsburger:2013b,Aidelsburger:2013,Ketterle:2013,Aidelsburger:2014} to generate artificial fluxes and Chern bands in optical lattices. This general method uses superlattices  (or Wannier-Stark ladders) with large static energy offsets $\Delta$, which inhibit the bare tunneling between neighboring sites, together with a resonant time-dependent modulation with characteristic frequency $\omega=\Delta/\hbar$; the latter restores the hopping in a controlled manner, e.g., generating complex tunneling matrix elements.

\begin{figure}[t!]
\includegraphics[width=8cm]{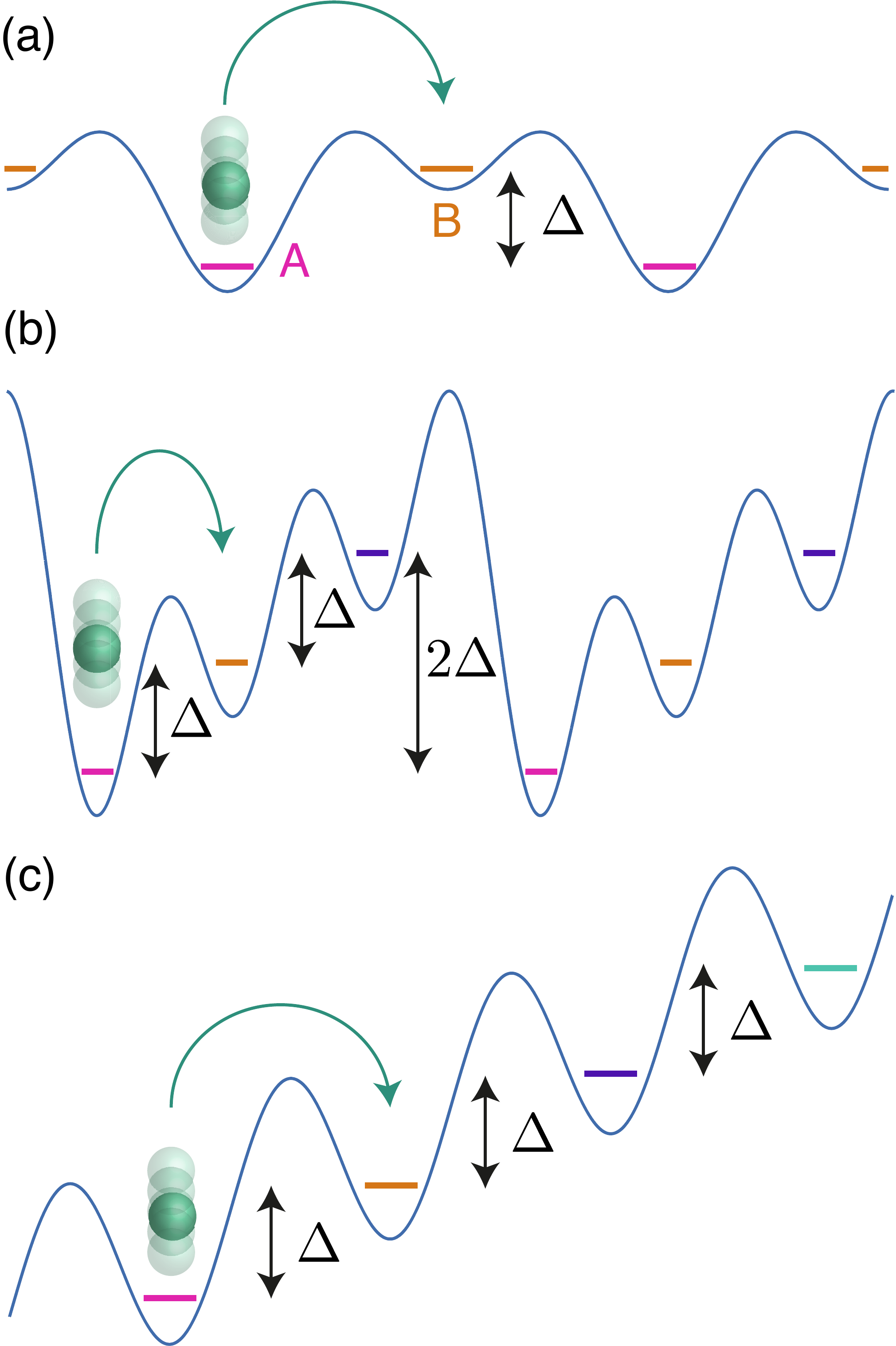}
\vspace{-0.cm} \caption{Superlattices and resonant modulations. (a) The two-site superlattice ($\alpha=0,1$), (b) the three-site superlattice ($\alpha=0,1,2$), and (c) the Wannier-Stark ladder ($\alpha=0,1,2,3, \dots$). Colors refer to the different sectors $\alpha \in \mathbb{Z}$. The cases (a) and (c) involve a single-harmonic modulation, with frequency $\omega = \Delta / \hbar$, see Section \ref{section_one_harmonics}. The case (b) involves a higher-order offset $2 \Delta$, requiring the use of a two-harmonic modulation to restore the tunneling over the entire lattice, see Section \ref{section_more}.}\label{Fig_Schema}
\end{figure}

We take the static Hamiltonian $\hat{H}^{(0)}$ to have
the form
\begin{equation}
\hat{H}^{(0)} = \sum_{\alpha\beta} \hat{P}^{\alpha} \hat{H}^{(0)}_{\alpha\beta} \hat{P}^\beta + \hbar\omega \sum_{\alpha} \alpha  \hat{P}^{\alpha}, \quad \alpha, \beta \in \mathbb{Z},
\label{eq:hoffset}
\end{equation}
where $\hat{P}^\alpha$ is a projection operator,
which divides the full Hilbert space into a set of
orthogonal sectors ($\hat{P}^\alpha\hat{P}^\beta = \delta_{\alpha\beta}\hat{P}^\alpha$) labelled by the integer $\alpha$, and $\sum_\alpha
\hat{P}^\alpha = \hat{\openone}$. The number of such sectors depends
on the problem of interest. A simple example is provided by a
tight-binding lattice Hamiltonian for a particle hopping on a superlattice
with two sites per unit cell, `A' and `B', with energy offsets
$0$ and $\hbar\omega$, respectively.  The full Hilbert space then
splits into states on `A' sites denoted by $\alpha = 0$, and those on
 `B' sites denoted $\alpha = 1$, see Fig.\ref{Fig_Schema} (a).  Such a case applies in the
experiments of Refs.~\cite{Aidelsburger:2011,Aidelsburger:2013b,Aidelsburger:2014}.  More generally,
the number of sectors $\alpha$ could be more than two, e.g. see Fig.\ref{Fig_Schema} (b) and Ref.\cite{Kennedy:2013}, and could
even be infinite, e.g. as in the Wannier-Stark
ladder, see Fig.\ref{Fig_Schema} (c) and Refs.~\cite{Aidelsburger:2013,Ketterle:2013}. A similar situation, of static energy offsets at multiples of $\hbar\omega$, has been studied by Hauke {\it et al.}\cite{Hauke:2012}. 

Using the notation of projectors, the time-dependent components in Eq. \eqref{eq:h} may be written as
\begin{eqnarray}
\hat H^{(j)} & = & \sum_{\alpha\beta}\hat{P}^\alpha \hat{H}_{\alpha\beta}^{(j)} \hat{P}^\beta  \quad  \text{ for } j \ne 0 .
\label{eq:vtproj}
\end{eqnarray}
 We stress that the present approach assumes that all divergent terms are present in the static Hamiltonian \eqref{eq:hoffset} only, and that they can be assembled in the form $ \hbar\omega \sum_{\alpha} \alpha  \hat{P}^{\alpha}$; in particular, we impose that the components $\hat{H}^{(j)}_{\alpha\beta}$ in Eqs. \eqref{eq:hoffset}-\eqref{eq:vtproj} do not contain any divergent terms.

We analyse this generalized situation, with Hamiltonian (\ref{eq:h})
formed from (\ref{eq:hoffset}) and (\ref{eq:vtproj}), by performing a
time-dependent unitary transformation
\be
|\psi\rangle \to |\psi'\rangle \!=\! \hat{R}(t) |\psi\rangle , \, \,  \hat{R}(t) \equiv
\exp\left[i\sum_\alpha \alpha \omega t \hat{P}^\alpha\right]  \!=\!\hat{R}(t\!+\!T).\label{unitary}
\ee
The new Hamiltonian, 
$\hat{\mathcal H} = \hat{R}\hat{H}\hat{R}^\dag - i\hbar \hat{R}\partial_t\hat{R}^\dag$,
can then be recast in the form \eqref{eq:h} studied in Ref. \cite{goldmandalibard}
\be
\hat{\mathcal H}(t)  = \sum_{j } \hat{\mathcal H}^{(j)} e^{ij\omega t}, \, \,  \hat{\mathcal H}^{(j)} = \sum_{\alpha\beta} \hat{P}^\alpha
\hat{H}_{\alpha\beta}^{(j-\alpha + \beta)} \hat{P}^\beta  , \label{gen_transf}
\ee
where calligraphic characters will hereafter be associated with the transformed frame. Importantly, the resulting time-dependent Hamiltonian $\hat{\mathcal H}(t)$ no longer contains diverging terms proportional to $\omega$. The new static term $\hat {\mathcal H}^{(0)}$ has contributions from the initial static elements $\hat{H}^{(0)}_{\alpha\alpha}$, which do not couple different sectors [e.g. onsite potentials or interactions], but also, from time-varying effects captured by $\hat{H}_{\alpha\beta}^{(j)}$, which do couple different sectors with $\beta - \alpha = j \ne 0$. Similarly, we note that the new time-dependent elements $\hat{\mathcal H}^{(j)}$, with $j\ne0$, can have contributions from the initial static terms $\hat{H}^{(0)}_{\alpha\beta}$ that couple different  sectors, $\alpha \ne \beta$ [e.g. tunneling terms].

Since $\hat{\mathcal H}(t)$ remains periodic in time, the methods of
Ref.~\cite{goldmandalibard} can be applied to understand its
effects as $\omega \to \infty$.  In particular, the time-evolution operator in Eq. \eqref{partition} now reads \cite{goldmandalibard}
\begin{align}
&\hat U(t;t_0) \!=\!\hat R^{\dagger} (t)e^{-i\hat{\mathcal K}(t)}e^{- \frac{i}{\hbar} (t - t_0)  \hat{\mathcal H}_{\rm eff}}e^{i\hat{\mathcal K}(t_0)} \hat R (t_0),\label{partition_2}\\
&\hat{\mathcal H}_{\rm eff}=\hat {\mathcal H}^{(0)} + \frac{1}{\hbar \omega } \sum_{j>0} \frac{1}{j}  \bigl [\hat {\mathcal H}^{(+j)} , \hat {\mathcal H}^{(-j)} \bigr ]    \!+\! \mathcal{O} (1/\omega^2) ,  \label{effective_ham_two} \\
&  \hat{\mathcal K} (t) \!=\! \frac{1}{i \hbar \omega}  \sum_{j>0} \frac{1}{j} \left [ \hat {\mathcal H}^{(+j)} e^{i j \omega t} \!-\! \hat {\mathcal H}^{(-j)} e^{-i j \omega t}     \right ]  \!+\! \mathcal{O} (1/\omega^2). \label{kick_two} 
\end{align} 
These expressions rely on a perturbative expansion in powers of $(1/\omega)$. In this work, we truncate the expansion to first order, noting that higher-order corrections are typically small in experimental situations \cite{Aidelsburger:2014};  expressions for higher-order terms are given in the Appendix, see also Refs.~\cite{goldmandalibard,Bukov:2014_review}.  Importantly, the convergence of the series in Eqs. \eqref{effective_ham_two}-\eqref{kick_two} relies on the fact that all the diverging terms contained in the time-periodic Hamiltonian \eqref{eq:h} have been removed by the unitary transformation in Eq. \eqref{unitary}. However, this is not necessarily the case in general \cite{goldmandalibard,Bukov:2014_review}, in particular, in situations where the strength of the time-modulation is also of order $\omega$. A generalization of the formalism handling additional (possibly time-dependent) diverging terms is presented in Section \ref{section_magnus}. Moreover, our analysis can readily be extended
to cases where the static offsets in (\ref{eq:hoffset}) are multiples
of an energy $\hbar\omega'$, with $\omega'$ any rational fraction of
the drive frequency $\omega$ ($\omega'/\omega \equiv N/M$ with $N$,
$M$ integer): the unitary transformation that removes $\omega'$ from
$\hat{H}^{(0)}$ again generates a time-dependent Hamiltonian $\hat{\mathcal H}(t)$ that is
periodic, now with frequency $\omega/M$. 


\subsection*{Advantages of the method}

Periodically-driven systems are often treated using  a stroboscopic analysis of the time-evolution: the state of the system $\vert \psi (t) \rangle$ is studied at specific times $t\!=\! N T$, where $N$ is an integer. According to the time-periodicity of the Hamiltonian $\hat H (t\!+\!T)\!=\!\hat H (t)$, the time-evolution operator can be written as $\hat U (t\!=\! N T) \!=\! \left [ U (T) \right ]^N $, where
\begin{align}
 \hat U (T)=\mathcal{T} e^{- \frac{i}{\hbar} \int_0^T \hat H (t) \text{d}t }   = e^{ - \frac{i}{\hbar} T \hat H_{\text{F}} }.\label{floquet}
\end{align}
Here $\mathcal{T}$ denotes the time-ordering, and  we introduced an  effective time-independent Hamiltonian $\hat H_{\text{F}}$.  The latter can be constructed perturbatively, for instance, using the Baker-Campbell-Hausdorff formula~\cite{goldmandalibard,Xu2013,Anderson2013} or the Magnus expansion~\cite{Maricq:1982,Bukov:2014_review}. The effective Hamiltonian $\hat H_{\text{F}}$ provides the effective band structure and the topological properties of the driven system \cite{Kitagawa:2010}. Importantly, by definition, this  stroboscopic analysis disregards any effects due to the micro-motion, which can lead to relevant effects in realistic situations \cite{goldmandalibard}; it also assumes that the modulation has been launched at a precise time [$t_0\!=\!0$ in Eq.~\eqref{floquet}]. One way to evaluate micro-motion effects in the stroboscopic picture is to treat the final (observation) time $t \!\ne\! NT$ as a random variable uniformly distributed within a driving period; the dynamics can then be numerically obtained through a special time-averaging over this independent random variable  \cite{Bukov:2014_review}. 

In contrast to the  stroboscopic analysis, the present method built on Eqs.~(\ref{unitary},\ref{partition_2}-\ref{kick_two}) allows for a complete description of the time-evolution, including the effects due to the initial phase of the driving and the micro-motion. Indeed, the time-evolution operator in Eq. \eqref{partition_2} can be systematically calculated, for any arbitrary initial driving time $t_0$ and final time $t \!\ne\! t_0 \!+\! NT$, through the evaluation of commutators [Eq.~\eqref{effective_ham_two}]. In particular, according to Eq. \eqref{partition_2}, the micro-motion is fully captured by the product of operators
\be
\hat R^{\dagger} (t) \exp (-i\hat{\mathcal K}(t)) \equiv e^{-i \hat{\mathcal{M}} (t)} ,\label{def_micro_operator}
\ee 
which can be explicitly calculated  using Eqs. (\ref{unitary},\ref{kick_two}). The right-hand side of Eq. \eqref{def_micro_operator} defines the \emph{micro-motion operator} $\hat{\mathcal{M}} (t)$, which will be explicitly computed and analyzed below.

We point out that the present  method can be readily applied to arbitrarily complicated time-dependent Hamiltonians. As for any perturbative approach, the present method is applicable in physical situations where the expansion in powers of $1/\omega$ rapidly converges, which is the case whenever the transformed Hamiltonian in Eq. \eqref{gen_transf} is regular. The next Section~\ref{section_magnus} discusses the applicability of the present method in situations where additional diverging features are present in the system, which typically occurs in a strong-driving regime  \cite{Bukov:2014_review}.

\section{Treating additional diverging terms: \\The strong-driving regime}\label{section_magnus}

In this Section, we generalize the formalism presented in Section \ref{section_general} to the case where the time-dependent Hamiltonian in Eqs. (\ref{eq:h},\ref{eq:hoffset},\ref{eq:vtproj}) contains additional diverging terms, see also Refs. \cite{goldmandalibard,Hauke:2012,Bukov:2014_review}. Here, we are interested in solving the Schrödinger equation for a time-periodic Hamiltonian of the general form
\be
\hat H(t) = \hat H_{\text{reg}}(t) + \hbar \omega \hat{O} (t),\label{eq:12}
\ee
where $\hat H_{\text{reg}}(t)$ and $\hat{O} (t)$ remain finite as $\omega \! \rightarrow\! \infty$, such that the second term explicitly diverges linearly in $\omega$.  Note that $\hat{O} (t)$ can contain a static part, as analyzed in the previous Section \ref{section_general} [Eq. \eqref{eq:hoffset}], as well as a time-dependent part representing ``strong driving". This strong-driving problem can be treated by generalizing the unitary transformation in Eq. \eqref{unitary}, so as to remove all diverging terms from the Hamiltonian $\hat H (t)$ in Eq.~\eqref{eq:12}, namely
\be
\vert \psi \rangle \rightarrow \vert \psi ' \rangle = \hat R (t) \vert \psi \rangle, \, \hat R (t)= \mathcal{T} \exp \left \{ i \omega \int_0^t \hat{O} (\tau) \text{d} \tau \right \},
\ee
which indeed leads to the transformed Hamiltonian 
\be
\hat{\mathcal{H}} (t)\!=\!\hat R(t)\hat H_{\text{reg}}\hat R^{\dagger} (t).\label{reg_transf}
\ee
This operation allows for significant progress in the resolution of the time-dependent system if the latter belongs to a class where $[\hat{O} (t),\hat{O} (t')]\!=\!0$ for all  times $t$ and $t'$; in other words, this operation is relevant for situations where $\hat R (t)$ and $\hat{\mathcal{H}} (t)$ can be computed explicitly. Considering such a class of systems, we impose that the operator $\hat R (t)$ should be time-periodic, 
\be
\hat R (t)= \exp \left \{ i \omega \int_0^t \hat{O} (\tau) \text{d} \tau \right \}  = \hat R (t+T),
\ee
so that the transformed Hamiltonian $\hat{\mathcal{H}} (t)$ in Eq.~\eqref{reg_transf} can be readily treated using the formalism presented in Section \ref{section_general} [Eqs. (\ref{partition_2}-\ref{kick_two})]. Note that this condition requires the time-average $(1/T)\int_0^T\hat{O}(\tau)\text{d}\tau$ to have eigenvalues that are integers. 

We are interested in treating systems with several diverging terms, e.g. $\hat{O} (t)=\hat{O}_1 (t) + \hat{O}_2 (t)$. In this framework it is useful to note the factorization rule $\hat R (t)=\hat R_1 (t) \hat R_2 (t)$, which is due to the commutativity $[\hat{O}_1,\hat{O}_2]=0$ of individual components at all times. This results in the transformed Hamiltonian
\be
\hat{\mathcal{H}} (t)=\hat R_2 (t)\hat R_1 (t)\hat H_{\text{reg}}\hat R_1^{\dagger} (t)\hat R_2^{\dagger} (t) .\label{transf_factorized}
\ee

For the sake of simplicity, we now apply this generalization of the formalism to a common situation, where the Hamiltonian includes a regular static term, a diverging static offset term and a strong resonant cosine modulation
\begin{align}
&\hat H (t) = \sum_{\alpha\beta} \hat{P}^{\alpha} \hat{H}^{(0)}_{\alpha\beta} \hat{P}^\beta + \hbar \omega \left ( \hat{O}_1+ \hat{O}_2 (t) \right ),\notag\\
&\hat{O}_1=\sum_{\alpha} \alpha  \hat{P}^{\alpha} , \quad \hat{O}_2 (t) = K_0 \cos (\omega t + \phi) \hat A , \label{quasi_general}
\end{align}
where the projectors $\hat P^{\alpha}$ and integers $\alpha$ have already been introduced in Eq. \eqref{eq:hoffset}. In order to fulfill the commutation conditions discussed above, the general operator $\hat A$ should necessarily commute with $\hat{O}_1$.  Generalizations to cases involving additional terms in Eq.~\eqref{quasi_general} can be treated along the same line, as long as the system satisfies the commutation conditions. 

First, we note that the unitary transformation related to the diverging static term $\hbar \omega \hat O_1$ has already been analyzed in Section \ref{section_general}; its associated operator is $\hat R_1 (t)= \exp\left[i\omega t \sum_\alpha \alpha \hat{P}^\alpha \right ]$, and the transformed Hamiltonian was given in Eq. \eqref{gen_transf}. The unitary transformation that removes the time-dependent term in Eq. \eqref{quasi_general} is written as
\be
\hat R_2 (t) = \exp \left [i K_0 \sin (\omega t + \phi) \hat A \right ] = \sum_{k = - \infty}^{\infty} \mathcal J_k (K_0 \hat A) e^{i k (\omega t + \phi)},\label{remove_drive}
\ee
where $\mathcal J_k$ denotes the Bessel functions of the first kind. The transformed Hamiltonian in Eq.\eqref{transf_factorized} is eventually given by
\begin{align}
&\hat{\mathcal{H}} (t)= \sum_j \hat{\mathcal{H}}^{(j)} e^{i j \omega t} , \label{transf_bessel} \\
&\hat{\mathcal{H}}^{(j)} \!=\! \sum_{\alpha , \beta} \sum_{k=- \infty}^{\infty} \mathcal J_k (K_0 \hat A)\hat{P}^{\alpha} \hat{H}^{(0)}_{\alpha\beta} \hat{P}^\beta \mathcal J_{k-j + \alpha - \beta} (K_0 \hat A) e^{i \phi (j + \beta - \alpha)}.\notag
\end{align}
The transformed Hamiltonian $\hat{\mathcal{H}} (t)$ in Eq. \eqref{transf_bessel} is time-periodic with period $T$, so that the time-evolution operator in Eq.\eqref{partition_2} can be constructed using the formalism described in Section \ref{section_general}, namely, 
\be
\hat U(t;t_0) \!=\!\hat R_2^{\dagger} (t)\hat R_1^{\dagger} (t)e^{-i\hat{\mathcal K}(t)}e^{- \frac{i}{\hbar} (t - t_0)  \hat{\mathcal H}_{\rm eff}}e^{i\hat{\mathcal K}(t_0)} \hat R_1 (t_0)\hat R_2 (t_0),\label{full_time_strong}
\ee
where the effective Hamiltonian $\hat{\mathcal H}_{\rm eff}$ and kick operators are given by the series in Eqs. \eqref{effective_ham_two}-\eqref{kick_two}. In particular, the micro-motion is now described by the product of operators
\be
e^{-i \hat{\mathcal{M}} (t)} = \!\hat R_2^{\dagger} (t) \hat R_1^{\dagger} (t) \, e^{-i\hat{\mathcal K}(t)}.\label{micro-motion_operator}
\ee
Importantly, the micro-motion depends on the time-modulation $\hat O_2 (t)$, but also on the static offset terms $\hat O_1$. This important aspect will be analyzed later in this work, based on concrete  examples.\\

\subsection*{Illustration for a strongly-driven two-level system}

We now illustrate this approach dedicated to strongly driven systems, by considering a simple example: two coupled levels, $\vert 0 \rangle$ and $\vert 1 \rangle$, separated by a very large energy offset $\Delta$, and subjected to a resonant driving with frequency $\omega = \Delta / \hbar$. We write the corresponding time-periodic Hamiltonian in the form \eqref{quasi_general},
\begin{align}
&\hat H(t) = \hat H_{\text{reg}} + \hbar \omega \left ( \hat{O}_1 +  \hat{O}_2 (t) \right ),\label{simple_example} \\
&\hat H_{\text{reg}}\!=\!\vert 0 \rangle \langle 1 \vert +  \vert 1 \rangle \langle 0 \vert , \, \, \hat{O}_1=\hat P^1 , \,\, \hat{O}_2 (t)= K_0  \cos (\omega t + \phi)  \hat P^0,\notag
 \label{simple_ex}
\end{align}
where the projectors onto the two levels $\alpha\!=\!0,1$ are given by $\hat P^{\alpha}\!=\!\vert \alpha \rangle \langle \alpha \vert$.  Applying the unitary transformations $\hat R_{1,2} (t)$ defined above [Eq. \eqref{remove_drive}], we obtain the transformed Hamiltonian $\hat{\mathcal{H}} (t)$ in Eq. \eqref{transf_bessel}, with the explicit Fourier components
\be
\hat{\mathcal{H}}^{(j)} \!= \mathcal{J}_{j+1} (K_0) \vert 0 \rangle \langle 1 \vert e^{i \phi (j +1)} + \mathcal{J}_{1-j} (K_0)  \vert 1 \rangle \!\langle 0 \vert e^{i \phi (j -1)},\label{fourier_bessel}
\ee
where we used the fact that
\be
 \mathcal{J}_{k} (K_0 \hat P^0) =  \mathcal{J}_{k} (K_0) \hat P^0 \!+\! \mathcal{J}_{k} (0) \hat P^1= \mathcal{J}_{k} (K_0) \hat P^0 + \delta_{k,0} \hat P^1.\notag
\ee
From the transformed Hamiltonian in Eq. \eqref{fourier_bessel}, one readily derives the effective Hamiltonian using Eq. \eqref{effective_ham_two}. To lowest order, this yields
\be
\hat{\mathcal H}_{\rm eff} \approx \hat{\mathcal{H}}^{(0)} =  \mathcal{J}_{1} (K_0) e^{i \phi} \, \vert 0 \rangle \langle 1 \vert  + \text{h.c.} \label{effective_bessel_one}
\ee
One recovers that the modulation essentially restores the coupling, and that the new coupling matrix elements are renormalized by a Bessel function of the first kind \cite{Eckardt:2007epl,Kolovsky:2011,Bermudez:2011prl}. Note that these matrix elements also acquire a complex phase factor, which is related to the phase of the modulation $\phi$; these induced complex phase-factors constitute the basis for generating artificial magnetic fields in modulated superlattices [Section \ref{section_one_harmonics}]. We also emphasize that the effective Hamiltonian in Eq. \eqref{effective_bessel_one} can be written as the time-average of the transformed Hamiltonian, $\hat{\mathcal H}_{\rm eff} \!\approx\! (1/T)\int_0^T \hat{\mathcal{H}} (t) \text{d}t$; this indicates that, at the lowest order of the perturbative treatment, the effective Hamiltonian \eqref{effective_bessel_one} is strictly equivalent to the one that would have been derived using a Magnus-expansion approach \cite{Hauke:2012,Bukov:2014_review}. Finally, we point out that the effective Hamiltonian in Eq. \eqref{effective_bessel_one} can be obtained in a similar manner for the case where the static energy offset in Eq.~\eqref{simple_example} is given by $N \hbar \omega \hat P^1$, with $N \!\in\! \mathbb{Z}$. In this case, the effective tunneling matrix elements in Eq. \eqref{effective_bessel_one} are found to be replaced by $\mathcal{J}_N (K_0) \exp (i \phi N)$.

The full time-evolution operator in Eq. \eqref{full_time_strong} is eventually obtained through the calculation of the kick operator $\hat{\mathcal{K}} (t)$ defined in Eq. \eqref{kick_two}. Using Eqs. (\ref{kick_two},\ref{fourier_bessel}), we find
\begin{align}
&\hat{\mathcal{K}} (t) \approx \frac{1}{i \hbar \omega} \sum_{j \ne 0} \frac{1}{j}  \mathcal{J}_{j+1} (K_0)  e^{i j (\omega t + \phi)} e^{i \phi} \,  \vert 0 \rangle \langle 1 \vert - \text{h.c. } \label{kick_bessel}\\
& \approx  \frac{1}{i \hbar \omega} \left \{ \vert 1 \rangle \langle 0 \vert  \left ( \mathcal{J}_{0} (K_0)e^{i \omega t} - \frac{\mathcal{J}_{1} (K_0)}{2}e^{i (2 \omega t + \phi)}  \dots    \right ) - \text{h.c. } \right \} . \notag
\end{align}
In particular, the micro-motion operator in Eq. \eqref{micro-motion_operator} is found to be well approximated by
\be
\hat{\mathcal{M}} (t) \approx  K_0 \sin (\omega t + \phi) \hat P^0   + \omega t \hat P^1 .\label{micro_approx}
\ee 

We conclude this paragraph by analyzing the weak-driving regime  of the system, i.e. $K_0 \ll 1$. In this case, the effective Hamiltonian \eqref{effective_bessel_one} and kick operators \eqref{kick_bessel} are now well approximated by
\be
\hat{\mathcal H}_{\rm eff} \approx \frac{K_0}{2} \vert 0 \rangle \langle 1 \vert e^{i \phi} + \text{h.c.}, \, \, \, \, \, 
\hat{\mathcal{K}} (t) \approx \frac{e^{i \omega t}}{i \hbar \omega}  \vert 1 \rangle \langle 0 \vert   + \text{h.c.},\notag
\ee 
whereas the micro-motion operator \eqref{micro-motion_operator} is still well approximated by Eq. \eqref{micro_approx}. We emphasize that these weak-driving results could have been equally obtained by directly applying the formalism of Section \ref{section_general}. 

In the following Sections, we will implicitly assume that the strength of the time-modulation is sufficiently weak so that the formalism of Section \ref{section_general} directly applies. However, we point out that the strong-driving regime of the time-modulated systems presented below can be  treated according to the general method discussed in this Section [see Section \ref{section_strong_mod}]. 


\section{Two-site dynamic superlattices: a momentum-space approach} \label{section_nigel} 

In this Section, we illustrate the formalism of Section \ref{section_general} by analyzing the general features of modulated two-site lattice systems, treated in a tight-binding description. Disregarding the actual geometry of the lattice, we consider models displaying two types of lattice sites (A and B, labelled by the sector index $\alpha =0,1$), and which are subjected to a local (on-site) time-dependent potential with a single harmonic. To simplify the presentation in this Section, we replace $\hat{P}^\alpha \hat{O}_{\alpha\beta}\hat{P}^\beta \to  \hat{O}_{\alpha\beta}$, taking the projectors to be implied by the subscripts. Using these shorter notations, the two-site static Hamiltonian  takes the general form 
\begin{equation}
\label{eq:2site_staticham}
\hat{H}^{(0)} = \sum_{\alpha , \beta =0,1}\hat{H}^{(0)}_{\alpha \beta} + \hat{P}^{1} \hbar\omega .
\end{equation}
Assuming that the time-dependent potential is only constituted of on-site operators, we specifically write the time-dependent components \eqref{eq:vtproj} with sector-diagonal entries
\be
\hat{H}^{(\pm 1)} = \hat{H}^{(\pm 1)}_{00} +  \hat{H}^{(\pm 1)}_{11}, \quad  \left [ \hat{H}^{(+1)}_{\alpha \alpha} , \hat{H}^{(- 1)}_{\alpha \alpha} \right ] =0 . \label{eq:2site_dynamicham}
\ee

This simple and general setting \eqref{eq:2site_staticham}-\eqref{eq:2site_dynamicham} may lead to interesting flux configurations and band properties \cite{Baur:2014ux}, as will be more specifically illustrated in Sections \ref{section_two_site}-\ref{section_honey}.

In order to evaluate the effective Hamiltonian \eqref{effective_ham_two} and kick operator \eqref{kick_two}, we perform the transformation (\ref{unitary}).  Since the corresponding unitary operator $\hat R (t)=\exp [ i  \omega t \hat{P}^{1}]$ only contains on-site terms, it commutes with all the terms in the time-dependent Hamiltonian, Eqs. \eqref{eq:2site_staticham}-\eqref{eq:2site_dynamicham}, except with the static inter-sector tunneling terms $\hat{H}^{(0)}_{10}$ and $\hat{H}^{(0)}_{01}$. Hence, appart from removing the diverging term in Eq. \eqref{eq:2site_staticham}, the only effect of the transformation (\ref{unitary}) is to make these tunneling terms time-dependent. This gives the modified Hamiltonian \eqref{gen_transf} with
\begin{align}
&\hat{\mathcal H}^{(0)}   =   \hat{H}^{(0)}_{00} + \hat{H}^{(0)}_{11} , \label{transf_two_site}\\
&\hat{\mathcal H}^{(1)}   =   \hat{H}^{(0)}_{10} + \hat{H}^{(1)}_{00}  + \hat{H}^{(1)}_{11}, \, \, \, \,  \hat{\mathcal H}^{(-1)}   =   \hat{H}^{(0)}_{01} + \hat{H}^{(-1)}_{00}  + \hat{H}^{(-1)}_{11}\, \notag.
\end{align}

\subsection{The effective Hamiltonian}

Taking the properties of the projection operators into account, the first-order effective Hamiltonian (\ref{effective_ham_two})  
 takes the form
\begin{align}
&\hat{\mathcal H}_{\rm eff}   =   \hat{H}^{(0)}_{00} + 
\hat{H}^{(0)}_{11}   + \frac{1}{\hbar\omega} \left[\hat{H}^{(0)}_{10}, \hat{H}^{(0)}_{01}\right] 
\label{eq:2site1} \\ 
&\!+\! \frac{1}{\hbar\omega} \bigl \{ \hat{H}^{(0)}_{10}\hat{H}^{(-1)}_{00}  \!+\! \hat{H}^{(1)}_{00} \hat{H}^{(0)}_{01}
\!-\! \hat{H}^{(-1)}_{11} \hat{H}^{(0)}_{10} \!-\! \hat{H}^{(0)}_{01}\hat{H}^{(1)}_{11} \bigr \}.
\label{eq:2site2}
\end{align}
The terms appearing in the first line (\ref{eq:2site1}) arise from the static
Hamiltonian (\ref{eq:2site_staticham}) with the inter-sector couplings treated within second order perturbation theory.
The terms in the second  line  (\ref{eq:2site2}) describe the restoration of couplings between A and B sectors via resonant modulations (\ref{eq:2site_dynamicham}).

We now make the additional assumption that the static system has spatial symmetries that lead to a conserved quasi momentum ${\bm k}$ and only {\it two} energy bands. This requires the corresponding Hamiltonian $\hat{H}^{(0)}$ to have a unit cell containing only one A and one B site. The wavevector ${\bm k}$ is then conserved (up to reciprocal lattice vectors) by all the terms in line (\ref{eq:2site1}). However, the time-modulation components $\hat{H}^{\pm 1}$ can have lower spatial symmetry: specifically, it can be that  $\hat{H}^{(-1)}_{\alpha\alpha}$ couples a state with wave vector ${\bm k}$    to a  state with wave vector ${\bm k}'= {\bm k}+{\bm q}$; such modulations that both transfer energy $\hbar \omega$ and momentum $\hbar \bs{q}$ to the system are crucial in the context of artificial gauge fields for cold-atoms \cite{Dalibard2011,Goldman:2013review}. Provided there is only one such wave vector  ${\bm q}$ (or wave vectors that differ from ${\bm q}$ by a reciprocal lattice vector), the quasi momentum ${\bm k}$ is also conserved by the inter-sector terms (\ref{eq:2site2}) of the effective Hamiltonian. That is,  the effective Hamiltonian only couples the state  with wave vector ${\bm k}$ on the A sites, $\vert 0,{\bm k}\rangle$, to  the state with wave vector ${\bm k}'={\bm k}+{\bm q}$ on the B sites,  $\vert 1,{\bm k}'\rangle$. In this two-state basis, $\{ \vert 0,{\bm k}\rangle ,  \vert 1,{\bm k}'\rangle \}$, the effective Hamiltonian \eqref{eq:2site1}-\eqref{eq:2site2} takes the form
\begin{align}
&\mathcal H^{\rm eff} (\bm k )  =  
\left(\begin{array}{cc}
\epsilon_{0}({\bm k})& v_{01} ({\bm k}) \\ v_{10}({\bm k}) & \epsilon_{1}({\bm k}')
\end{array}\right)  , \label{eq:2site_heff} \\
& \epsilon_{0}({\bm k})\equiv \langle 0,{\bm k}| \hat{H}^{(0)}_{00}-\frac{1}{\hbar\omega}\hat{H}^{(0)}_{01}\hat{H}^{(0)}_{10}|0,{\bm k}\rangle , \notag \\
&\epsilon_{1}({\bm k}')\equiv \langle 1,{\bm k}'| \hat{H}^{(0)}_{11}+\frac{1}{\hbar\omega}\hat{H}^{(0)}_{10}\hat{H}^{(0)}_{01}|1,{\bm k}'\rangle , \notag \\
&v_{10}({\bm k})\equiv  \frac{1}{\hbar\omega}\langle 1,{\bm k}'|\hat{H}^{(0)}_{10}\hat{H}^{(-1)}_{00}-\hat{H}^{(-1)}_{11}\hat{H}^{(0)}_{10}|0,{\bm k}\rangle=v^*_{01}({\bm k}).\notag
\end{align}
Here $\epsilon_{0}({\bm k})$ and $\epsilon_{1}({\bm k}')$ are the dispersions for particles of wave vectors ${\bm k} $ and ${\bm k}'$ moving on the decoupled A  and B sites, as described by the terms in (\ref{eq:2site1}), and $v_{10}({\bm k})$ encodes the couplings between the A and B sites through the terms  (\ref{eq:2site2}). The energy spectrum and topological properties associated with the effective Hamiltonian $\hat{\mathcal H}_{\rm eff}$ can then be directly deduced from the functions $ \epsilon_{0,1}$ and  $v_{10}$ in Eq. \eqref{eq:2site_heff}, as will be illustrated in Section \ref{section_honey} using a specific model.

\subsection{The kick operator} \label{section_kick_two_site}

 For such models, the micro-motion undergone within each period of the driving [Eq. \eqref{def_micro_operator}] is essentially due to the kick operator \eqref{kick_two}, whose leading terms  are given by
\begin{equation}
\hat{\mathcal K} (t) 
= \frac{1}{i\hbar\omega}\left( \hat{H}^{(0)}_{10} + \hat{H}^{(1)}_{00}  + \hat{H}^{(1)}_{11}\right)e^{i\omega t} + \mbox{h.c.} ,
\end{equation}
where we used Eq. \eqref{transf_two_site}. For ${\bm q}= 0$, this couples the two basis states $|0,{\bm k}\rangle$, $|1,{\bm k}\rangle$.
For ${\bm q}\neq 0$,  the kick produced by the operator $\exp \left [-i \hat{\mathcal K} (t) \right ]$ in Eq. \eqref{partition_2} can couple the set of states  $\{ |0,{\bm k}+ n{\bm q}\rangle, |1,{\bm k}+ m{\bm q}\rangle\}$ with $n$, $m$ any  integers in the range $0\leq n,m\leq n_{\rm max}-1$, and $n_{\rm max}$ defined by the condition that $n_{\rm max} {\bm q}$ is a reciprocal lattice vector.   In this case, the micro-motion will involve oscillations between plane-wave states at these $n_{\rm max}$ different wave vectors ${\bm k}+n{\bm q}$. As will be discussed below in specific examples, this leads to an oscillation in the amplitudes of discrete peaks in the momentum distribution, as revealed by time-of-flight images in cold-atom setups; we shall show an example in Section \ref{section_two_site}, leading to staggered flux pattern, where $n_{\rm max}=4$, and thus displaying four peaks in the expansion images, see Fig. \ref{Fig_Micro}.

\section{Modulated square super-lattices}\label{section_one_harmonics}

We now apply the formalism to the case of general square superlattices, which are modulated in time in a resonant manner in view of realizing non-trivial flux patterns. 

\subsection{The time-dependent Hamiltonian}

Considering a single-band tight-binding approximation, the static Hamiltonian $\hat H^{(0)}$ is taken in the general second-quantized form
\begin{align}
\hat{H}^{(0)}= \hat T_x + \hat T_y +  \hat S + \hat U_{\text{onsite}}, \quad \hat S= \Delta \sum_{m,n} s(m)  \hat n_{m,n} . \label{statone}
\end{align}
The static Hamiltonian includes the nearest-neighbor hopping terms 
\begin{align}
&\hat T_x= -J_x \sum_{m,n} \hat a_{m+1,n}^{\dagger} \hat a_{m,n} +  \text{h.c.} ,\notag \\
&\hat T_y= -J_y \sum_{m,n} \hat a_{m,n+1}^{\dagger} \hat a_{m,n} + \text{h.c.} ,\notag 
\end{align}
where $J_{x,y}$ denote the hopping matrix elements, $\hat a_{m,n}^{\dagger}$ creates a particle at lattice site $\bs x = (ma,na)$, $a$ is the lattice spacing, $(m,n)$ are integers. The number operator in Eq. \eqref{statone} is defined as $\hat n_{m,n} = \hat a_{m,n}^{\dagger} \hat a_{m,n}$. The Hamiltonian in Eq. \eqref{statone} also includes a ``superlattice" term $\hat S$, which creates energy offsets between lattice sites along the $x$ direction, the spatial modulation being described by the function $s(m)$. The energy $\Delta \gg J_x$ is large so that the bare tunneling is potentially inhibited along the $x$ direction, depending on the spatial modulation $s(m)$. For instance, the two-site superlattice potential [Fig. \ref{Fig_Schema} (a)] corresponds to the case $s(m)=\nicefrac{1}{2}(-1)^m$, whereas the Wannier-Stark ladder [Fig. \ref{Fig_Schema} (c)] corresponds to the case $s(m)=m$. Importantly, the superlattice function $s(m)$ only depends on the $x$ coordinate, which will simplify the following analysis. In Eq. \eqref{statone}, all additional static potentials (e.g. confinement),  and onsite inter-particle-interaction terms are assembled in $\hat U_{\text{onsite}}$:  in contrast to the tunneling terms, onsite potentials do not couple different sectors, and thus they commute with the unitary transformation \eqref{unitary}, see Eq. \eqref{transf_hof} below. Hence, they can be directly included in the effective Hamiltonian $\hat{\mathcal H}_{\rm eff}$ at the zero-th order level, $\hat {\mathcal H}^{(0)} \rightarrow  \hat {\mathcal H}^{(0)} + \hat U_{\text{onsite}}$, see Eqs. \eqref{gen_transf} and \eqref{effective_ham_two}; note also that these static onsite terms do not enter the kick operator in Eq. \eqref{kick_two}.

For the sake of simplicity, we will assume that $s(m+1) - s(m) = \pm 1$, in which case a single-harmonic modulation is sufficient to restore the tunneling. The time-dependent Hamiltonian is thus written in the form (\ref{eq:h},\ref{eq:hoffset},\ref{eq:vtproj}) \cite{footnote},
\begin{align}
&\hat H (t)= \hat H^{(0)} + \hat V (t) =  \hat H^{(0)} + \hat H^{(+1)} e^{i \omega t} + \hat H^{(-1)} e^{-i \omega t}, \notag \\
&\hat H^{(+1)}= \kappa \sum_{m,n} \hat n_{m,n} v (m,n) = \left [ \hat H^{(-1)}  \right]^{\dagger} , \quad \omega = \Delta / \hbar ,  \label{mod_def}
\end{align}
where the latter equality corresponds to the resonance condition, and where the static component $\hat H^{(0)}$ is given in Eq. \eqref{statone}. We stress that any small detuning from the resonance $\omega = \Delta / \hbar$ can be incorporated in the formalism, by adding a weak potential $\delta \sum_{m,n} s(m)  \hat n_{m,n} $ in the on-site operator $\hat U_{\text{onsite}}$ introduced above. Note also that we have included a spatial dependence $v(m,n)$ in the time-modulation $\hat V (t)$, which will play an important role in the following \cite{Kolovsky:2011,Creffield:2013gp,Aidelsburger:2013,Ketterle:2013}. The more general situation where the superlattice function $s(m)$ creates  ``high-order" energy offsets [$ 2\Delta , 3 \Delta , \dots$] will be treated in Section \ref{section_more}.

\subsection{The unitary transformation}

The unitary transformation in Eq. \eqref{unitary} takes an explicit form in terms of the superlattice operator $\hat S$,
\begin{align}
&\vert \psi ' \rangle = \hat R(t) \vert \psi \rangle= \exp \left (\! i  t \hat S / \hbar \right ) \vert \psi \rangle . \label{transf_hof} 
\end{align}
The transformed Hamiltonian reads
\begin{align}
\hat {\mathcal H} (t) = \hat{\mathcal{T}}_x + \hat T_y + \hat U_{\text{onsite}} + \hat V (t), \label{ch_frame}
\end{align}
where the modified tunneling term along the $x$ direction reads
\be
\hat{\mathcal T}_x = -J_x \sum_{m,n} \hat a_{m+1,n}^{\dagger} \hat a_{m,n} e^{i \omega t \left [ s(m+1) - s(m)   \right ]} + \text{h.c.} \label{ch_frame_2}
\ee
Only the tunneling term along the $x$ direction is affected by the unitary transformation \eqref{transf_hof}, since $[\hat T_y, \hat S]=0$, in the present case where the superlattice function $s=s(m)$ does not depend on the $y$ coordinate. Note also that the onsite operators $\hat U_{\text{onsite}}$ and $\hat V(t)$ are not affected by the transformation, as they also commute with $\hat S$.\\

As already announced above, we simplify the discussion by constraining the superlattice function,
\be
s(m+1) - s(m)  = \pm 1 \equiv \delta_s (m) . \label{def_m}
\ee
The more general case $\delta_s (m) \in \mathbb{Z}$ will be treated in Section \ref{section_more}. We find useful to label the sites according to the notation $m=m^{\pm}$ defined as
\be
\delta_s (m^+) = +1 , \qquad \delta_s (m^-) = -1, \label{def_m_2}
\ee
which classifies the sites along $x$ in terms of their nearest-neighbour offset [$\pm \Delta$], see Fig. \ref{Fig_Plaquette} (a). Using this notation, the time-dependent Hamiltonian $\hat {\mathcal H} (t)$ in Eqs. \eqref{ch_frame}-\eqref{ch_frame_2} can be written in the form \eqref{gen_transf}, 
\begin{align}
&\hat {\mathcal H} (t) = \hat {\mathcal H}^{(0)} +\hat {\mathcal H}^{(+1)} e^{i \omega t} + \hat {\mathcal H}^{(-1)} e^{-i \omega t} ,\label{ham_dec}\\
& \hat {\mathcal H}^{(0)} =\hat T_y + \hat U_{\text{onsite}}, \notag \\
&\hat {\mathcal H}^{(+1)} = \kappa \sum_{m,n} \hat n_{m,n}\,  v (m,n) \notag \\
&\qquad \quad - J_x \left \{ \sum_{m^+,n} \hat a_{m+1,n}^{\dagger} \hat a_{m,n} + \sum_{m^-,n} \hat a_{m,n}^{\dagger} \hat a_{m+1,n}\right \} = [\hat {\mathcal H}^{(-1)}]^{\dagger}. \notag
\end{align}
Note that the term $ \hat {\mathcal H}^{(0)} $ does not have any contribution from the time-dependent components $\hat H^{(\pm 1)}$. In contrast, the terms $\hat {\mathcal H}^{(\pm 1)}$ have contributions  from the static $\hat H^{(0)}$ and non-static $\hat H^{(\pm 1)}$ elements.

\subsection{The effective Hamiltonian and flux patterns}\label{section_effective}

We now compute the first-order contributions to the effective Hamiltonian $\hat{\mathcal H}_{\text{eff}}$ in Eq. \eqref{effective_ham_two}, using Eq. \eqref{ham_dec}:
\begin{widetext}
\begin{align}
\frac{1}{\hbar \omega } [\hat {\mathcal H}^{(+1)}, \hat {\mathcal H}^{(-1)}]= - \frac{J_x \kappa}{\hbar \omega} 
\Biggl \{ &\sum_{m^+,n} \hat a_{m+1,n}^{\dagger} \hat a_{m,n} \gamma (m,n)  - \sum_{m^-,n} \hat a_{m+1,n}^{\dagger} \hat a_{m,n} \gamma^* (m,n) + \text{h.c.} \Biggr \}, \quad \gamma (m,n) = v^*(m,n) - v^*(m+1,n),\label{general_restored}
\end{align}
which shows that tunneling can be restored between neighboring sites, $(m,n) \leftrightarrow (m \pm 1,n)$, if and only if the time-modulation \eqref{mod_def} generates a differential shaking $v(m,n) \ne v(m \pm 1,n)$. This also suggests that tailoring the time-modulation $\hat V (t)$ allows to address different links independently, as recently implemented in Ref.~\cite{Aidelsburger:2014}, see Section \ref{section_two_site}.
\end{widetext}

\begin{figure}[b!]
\includegraphics[width=9.cm]{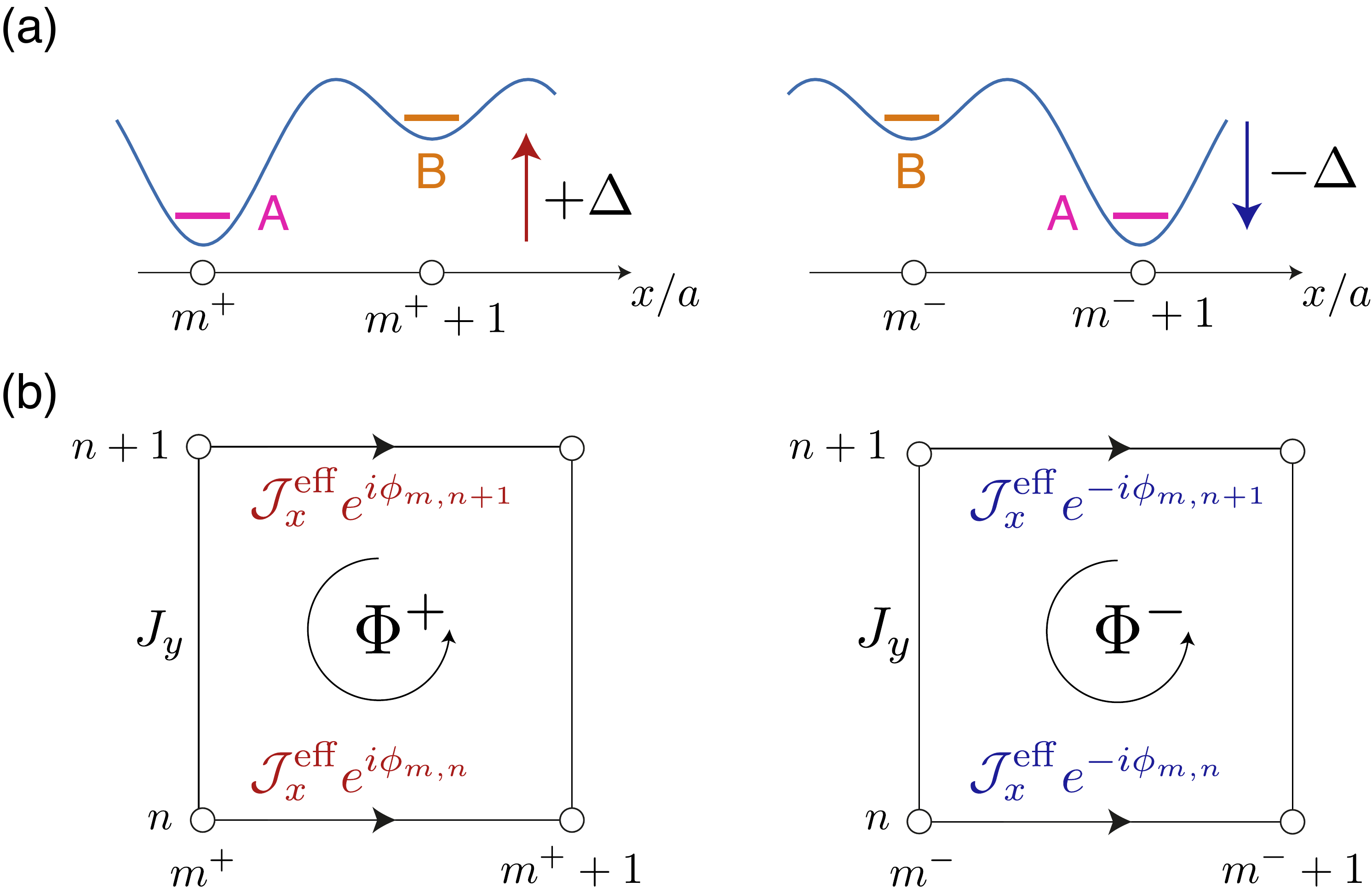}
\vspace{-0.cm} \caption{(a) Labeling of sites $m=m^{\pm}$ according to the energy offset $\pm \Delta$,  see Eqs. \eqref{def_m}-\eqref{def_m_2}. (b) The corresponding plaquettes $\square^{\pm}$, with Peierls phase-factors $\exp (\pm i \phi_{m,n})$ along the $x$ direction; the flux penetrating these two plaquettes are given by $2 \pi \Phi^+=\phi_{m^+,n} - \phi_{m^+ , n+1}$, and $ 2 \pi \Phi^-=\phi_{m^-,n+1} - \phi_{m^- , n}$, respectively. The phases $\phi_{m,n}$, defined in Eq. \eqref{uniform_hypo}, are established by the time-periodic modulation \eqref{mod_def}.}\label{Fig_Plaquette}
\end{figure}

We now make the assumption that the modulation can be designed so as to verify
\be
\gamma (m,n) = v^*(m,n) - v^*(m+1,n) = \rho \, e^{i \phi_{m,n}}, \label{uniform_hypo}
\ee
so that the restored tunneling amplitude $\sim \rho$ is uniform over the lattice ~\cite{Aidelsburger:2011,Aidelsburger:2013b,Aidelsburger:2013,Ketterle:2013,Aidelsburger:2014}.  This condition will be satisfied in specific examples that we provide below. The first-order effective Hamiltonian finally takes the form
\begin{align}
&\hat{\mathcal H}_{\rm eff}= \hat T_y + \hat U_{\text{onsite}} \label{ham_stag_1} \\
&\!-\! \mathcal{J}_x^{\text{eff}} \Biggl \{ \sum_{m^+,n}   \hat a_{m+1,n}^{\dagger} \hat a_{m,n} e^{i \phi_{m,n}} \!-\! \sum_{m^-,n}   \hat a_{m+1,n}^{\dagger} \hat a_{m,n} e^{-i  \phi_{m,n} } \!+\! \text{h.c.} \Biggr \} ,  \notag
\end{align}
where the induced tunneling amplitude is given by 
\be
\mathcal{J}_x^{\text{eff}}= J_x \kappa \rho / \hbar \omega ,\label{induced_ampli}
\ee
see Fig. \ref{Fig_Plaquette} (b). The latter result \eqref{ham_stag_1} shows how Peierls phase-factors $\phi_{m,n}$ are generated by the resonant modulation; their expressions are directly dictated by the modulation functions $v(m,n)$, defined in Eq. \eqref{mod_def}, through Eq. \eqref{uniform_hypo}. Second-order corrections to the effective Hamiltonian \eqref{general_restored}-\eqref{ham_stag_1}, which lead to a modification of the hopping term along the $y$ direction, are discussed in Appendix.  We point out that the derivation of the effective Hamiltonian in Eqs.\eqref{ham_stag_1}-\eqref{induced_ampli} assumes that the strength of the time-modulation is sufficiently weak,  $\kappa\! \ll\! \hbar \omega$, see Sections \ref{section_general} and \ref{section_magnus}. A generalization of this result, valid in the strong-driving regime $\kappa\! \sim\! \hbar \omega$, is presented in Section \ref{section_strong_mod}, based on the method introduced in Section \ref{section_magnus}.

We now evaluate the flux penetrating the diverse plaquettes. Using Eq. \eqref{ham_stag_1}, we find two types of plaquettes:
\begin{itemize}
\item plaquettes $\square^+$, characterized by sites of type $m^+$ at the left corners, see Eqs. \eqref{def_m}-\eqref{def_m_2} and Fig. \ref{Fig_Plaquette} (b). The corresponding flux is given by 
\be
2 \pi \Phi^+=\phi_{m^+,n} - \phi_{m^+ , n+1}. \label{flux_plus}
\ee
\item plaquettes $\square^-$, characterized by sites of type $m^-$ at the left corners, see Eqs. \eqref{def_m}-\eqref{def_m_2} and Fig. \ref{Fig_Plaquette} (b). The corresponding flux is given by 
\be
 2 \pi \Phi^-=\phi_{m^-,n+1} - \phi_{m^- , n}. \label{flux_minus}
 \ee
\end{itemize}
 Here, the space-dependence of the fluxes $\Phi^{\pm}=\Phi^{\pm}(m,n)$ is implicit. For phases of the form $\phi_{m,n}=c_x m + c_y n$, the fluxes  $\Phi^{\pm}=\mp c_y$ are necessarily constant along the $y$ direction, but potentially vary along the $x$ direction, depending on the superlattice function $s(m)$.

\subsection{Full-time evolution: the kick operator and the micro-motion}

The main contribution to the kick operator are given by [Eqs. \eqref{kick_two} and \eqref{ham_dec}]
\begin{align}
 \hat{\mathcal K} (t) &= \frac{1}{i \hbar \omega}  \left [ \hat {\mathcal H}^{(+1)} e^{i \omega t} - \hat {\mathcal H}^{(-1)} e^{-i \omega t}     \right ] \label{kick_three}\\
&\approx  \frac{2 \kappa}{\hbar \omega} \sum_{m,n} \hat n_{m,n}\, \vert v (m,n) \vert  \sin (\omega t + \theta_{m,n}),   \notag
\end{align}
where we assumed that $\kappa \gg J_x$ in the last equation, and where $\theta_{m,n} = \text{arg} [v (m,n)]$. 

We write the full-time-evolution operator \eqref{partition_2} as
\be
\hat U (t;t_0) = e^{-i \hat{\mathcal{M}} (t)} \hat U_{\text{eff}} (t; t_0) , \, \hat U_{\text{eff}} (t; t_0) =e^{- \frac{i}{\hbar} (t-t_0) \hat{\mathcal{H}}_{\rm eff}} e^{i \hat{\mathcal{M}} (t_0)} , \label{full_description}
\ee
where $\hat{\mathcal{H}}_{\rm eff}$ denotes the effective Hamiltonian in Eq. \eqref{ham_stag_1} and where the ``micro-motion" operator \eqref{def_micro_operator} is given by
\begin{align}
\hat{\mathcal {M}} (t)\!=\!\sum_{m,n} \hat n_{m,n} \Biggl \{  \omega t s (m) \!+\! \frac{2 \kappa}{\hbar \omega}  \vert v (m,n) \vert  \sin (\omega t \!+\! \theta_{m,n}) \Biggr \} .\label{micro_operator}
\end{align}
Hence, the micro-motion both depends on the superlattice (spatial) modulation $s(m)$, Eq. \eqref{statone}, as well as on the  function $v(m,n)$ characterizing the time-modulation, Eq. \eqref{mod_def}. Different micro-motions will be discussed below, based on specific examples. \\

\subsection{The strong-driving regime}\label{section_strong_mod}

The effective Hamiltonian and micro-motion operators can also be evaluated in the strong-driving regime, $\kappa \!\sim\! \hbar \omega$. Applying the method detailed in Section \ref{section_magnus} to the time-dependent Hamiltonian \eqref{statone}-\eqref{mod_def}, we find the effective Hamiltonian 
\begin{align}
\hat{\mathcal{H}}_{\text{eff}} \!\approx\! \hat U_{\text{onsite}} +  \sum_{m^{\pm},n} \, & (\pm) \mathcal{J}_x (m,n)e^{\pm i \phi_{m,n}}  \hat a_{m+1,n}^{\dagger} \hat a_{m,n} \notag \\ 
&+ \mathcal{J}_y (m,n) \hat a_{m,n+1}^{\dagger} \hat a_{m,n} + \text{h.c.}, \label{effective_harper}
\end{align}
where $\phi_{m,n}\!=\!\text{arg} [v(m+1,n) \!-\! v(m,n)]$, and where the effective hopping amplitudes are now given by the expressions
\begin{align}
&\mathcal{J}_x (m,n)= J_x \,  \mathcal{J}_1 \left ( 2K_0 \vert \delta_x v(m,n) \vert \right ), \label{amplitudes_strong} \\
&\mathcal{J}_y (m,n)= J_y \,  \mathcal{J}_0 \left ( 2K_0 \vert \delta_y v(m,n) \vert \right ), \qquad K_0=\kappa/\hbar \omega.\notag
\end{align}
Here $\delta_{x,y}$ denote finite-difference operations along the $x$ and $y$ directions, e.g. $\delta_{x} v(m,n)\!=\! v(m+1,n) \!-\! v(m,n)$. We note that the two types of Bessel functions in Eq.~\eqref{amplitudes_strong}, $\mathcal{J}_{1,0}$, are associated with the presence of static energy offsets $N \hbar \omega$, with $N=1,0$ along the $x$ and $y$ directions, respectively [see Section \ref{section_magnus}]. We readily verify that the effective Hamiltonian in Eqs. \eqref{effective_harper}-\eqref{amplitudes_strong} coincides with the result in Eq. \eqref{ham_stag_1} in the weak-driving limit $\kappa \!\ll\! \hbar \omega$, see also Appendix.   Additionally, the micro-motion operator \eqref{micro-motion_operator} is found to be still well described by Eq. \eqref{micro_operator} in the strong-driving regime.

\section{Applications: Flux rectification as a route towards Chern insulators}\label{section_uniform}

In this Section, we discuss several schemes leading to uniform flux per plaquette $\Phi=\Phi^{+}=\Phi^{-}$ over the entire lattice. This search is particularly motivated by the fact that this uniform-flux configuration -- also commonly known as the Harper-Hofstadter model \cite{Hofstadter:1976,Harper:1955,Azbel:1964} -- leads to topological Chern bands with interesting features. More specifically, for particular values of the flux $\Phi$, the lowest band of the bulk energy spectrum is associated with a non-zero Chern number $\nu_{\text{ch}} \ne 0$ and a large flatness ratio $f=\Delta_{\text{gap}}/W$, where $\Delta_{\text{gap}}$ denotes the spectral gap separating the lowest band from the upper bands, and where $W$ denotes the bandwidth. These topological properties, combined with a large flatness ratio, make such Chern bands good candidates for realizing fractional Chern insulators \cite{Bergholtz:2013,Parameswaran:2013}. 

For a square lattice with uniform flux $\Phi=1/4$, as realized in current experiments \cite{Aidelsburger:2013,Aidelsburger:2014}, the lowest band has a Chern number $\nu_{\text{ch}}=1$, with a flatness ratio $f \approx 7$ and a band gap $\Delta_{\text{gap}} \approx 1.53 J$, where $J/h \lesssim 100 \text{Hz}$ is the typical hopping amplitude over the lattice. Similar Chern bands with $\nu_{\text{ch}}=1$ are obtained for generic flux $\Phi=1/q$, with $q \in \mathbb{Z}$, see Refs. \cite{Thouless1982,Kohmoto:1989}. We note that while the flatness ratio $f$ increases with $q$, the band gap $\Delta_{\text{gap}}$ is maximized for $q=5$; for this optimized value, we find $\Delta_{\text{gap}} \approx 1.55 J$ and a large flatness ratio $f \approx 24$. The strong reduction of the band gap as $q$ is further increased is clearly revealed in the well-known Hofstadter butterfly \cite{Hofstadter:1976}.
More exotic configurations with $\vert \nu_{\text{ch}} \vert> 1$ can also be generated with flux of the form $\Phi=p/q$, where $p,q$ are integers: the Chern number of the lowest band satisfies the Diophantine equation \cite{Thouless1982,Kohmoto:1989}
\be
1=p \, \nu_{\text{ch}} + q \,  \sigma, \quad \vert \nu_{\text{ch}} \vert < q/2 , \quad \sigma \in \mathbb{Z} .
\ee 
For instance, setting $\Phi=4/9$, the Chern number of the lowest band is $\nu_{\text{ch}}=-2$, the flatness ratio is large $f \approx 13$ and the band gap $\Delta_{\text{gap}} \approx 0.3 J$ is still reasonable (i.e. of the order of current experimental temperatures). More complex settings leading to flat bands with arbitrary Chern numbers were discussed in Ref. \cite{Bergholtz:2013}. \\

\subsection{The main time-modulation, Peierls phase-factors and the micro-motion}\label{section_main_mod}

In this Section, we will mainly consider the simple time-periodic modulation used in the Munich and MIT experiments \cite{Aidelsburger:2011,Aidelsburger:2013, Ketterle:2013}, which is generated by a single pair of laser beams with frequency difference $\omega_1 - \omega_2=\omega$, and wave vector difference $\bs{k}_1 - \bs{k}_2=\bs{q}$. This configuration essentially provides a moving potential of the form
\begin{align}
\hat V (t)&\!=\! 2 \kappa \sum_{m,n} \hat n_{m,n} \cos \left ( \omega t \!+\! \bs{q} \cdot \bs{R}  \right )  , \,  \bs{R}/a \!=\! m \bs{1}_x + n \bs{1}_y ,\label{mod_munich_one} 
\end{align}
where $a$ is the lattice spacing. This corresponds to the on-site energy modulation in Eq.  \eqref{mod_def} with 
\be
v(m,n) = \exp (i q_x m a) \exp (i q_y n a).\label{mod_munich_one_bis}
\ee 
We note that this laser configuration also leads to a displacement of the lattice sites, and potentially to a deformation of the corresponding wells; these additional effects are assumed to be small compared to the on-site modulation in Eq. \eqref{mod_munich_one}, which is a reasonable assumption for the schemes realized so far \cite{Aidelsburger:2011,Aidelsburger:2013, Ketterle:2013,Aidelsburger:2014}. Besides, we remind that  a time-periodic displacement of the sites' position $\bs r_j (t)$ can be re-formulated in terms of an on-site energy modulation through a unitary (change-to-a-moving-frame) transformation \cite{goldmandalibard,Lignier:2007,Arimondo2012}.

We will describe the effects of the time-modulation \eqref{mod_munich_one}, based on different superlattice configurations with offsets $\Delta =\hbar \omega$. Using Eqs. \eqref{uniform_hypo} and \eqref{induced_ampli}, we can already state that the restored tunneling amplitude $\mathcal{J}_x^{\text{eff}}$ and Peierls phase-factors $\phi_{m,n}$ associated with the time-modulation \eqref{mod_munich_one}-\eqref{mod_munich_one_bis} are necessarily given by
\be
\mathcal{J}_x^{\text{eff}}=\frac{ \sqrt{2} J_x \kappa}{\hbar \omega} \biggl [1-\cos (a q_x) \biggr], \, \,  \, \, \phi_{m,n}=\frac{\pi}{2} - \bs{q} \cdot \bs{R} - \frac{a q_x}{2}.\label{induced_hop_peierls}
\ee
Moreover, the flux associated with the plaquettes $\square^{\pm}$, Eqs. \eqref{flux_plus}- \eqref{flux_minus}, are given by 
\be
2 \pi \Phi^{\pm}=\pm q_y a . \label{flux_general}
\ee
Summarizing, independently of the form given to the static superlattice potential [i.e. $s(m)$], the tunneling amplitude is controlled by $q_x$, while the effective flux per plaquette are dictated by $q_y$, which can be tuned, e.g., by changing the relative angle between the two laser beams. In particular, tunneling is restored whenever $q_x \ne (2 \pi/a) \times \text{integer}$, assuming that the frequency $\omega$ is resonant with the static energy offset $\Delta$.  Note that the case $q_x = (2 \pi/a) \times \text{integer}$ corresponds to an on-site modulation that is in phase between neighboring lattice sites, which precludes tunneling restoration. In the following, we will generally consider the case $q_x=q_y=\pi/2a$, as in the experiment \cite{Aidelsburger:2013}. This yields [Eqs. \eqref{induced_hop_peierls}-\eqref{flux_general}] 
\be
\mathcal{J}_x^{\text{eff}}=\frac{ \sqrt{2} J_x \kappa}{\hbar \omega} ,\quad \phi_{m,n}=-\frac{\pi}{2} (m\!+\!n \!-\!\frac{1}{2}), \quad \Phi^{\pm} = \pm 1/4 .\label{munich_phi}
\ee

Finally, considering the time-modulation \eqref{mod_munich_one}-\eqref{mod_munich_one_bis}, the micro-motion operator \eqref{micro_operator} takes the more explicit form
\begin{align}
\hat{\mathcal {M}} (t)=\sum_{m,n} \hat n_{m,n}\, \Biggl \{  \omega t s (m) + \frac{2 \kappa}{\hbar \omega}  \sin [ \bs{q}\cdot \bs{R} + \omega t ] \Biggr \} , \label{micro_motion_simple} 
\end{align}
where we remind that the function $s(m)$ is determined by the static superlattice potential $\hat S$.

\subsection{The Wannier-Stark ladder} \label{section_wannier_ladder}

In the case of the Wannier-Stark-ladder, Fig. \ref{Fig_Schema}(c), the superlattice function is given by $s(m)=m$ so that all the sites are of the type $m=m^+$, see Eqs. \eqref{def_m}-\eqref{def_m_2} and Fig. \ref{Fig_Plaquette}. The effective Hamiltonian is thus given by
\begin{align}
&\hat{\mathcal H}_{\rm eff}= \hat T_y + \hat U_{\text{onsite}} - \mathcal{J}_x^{\text{eff}}  \sum_{m^+,n}   \hat a_{m+1,n}^{\dagger} \hat a_{m,n} e^{i \phi_{m,n}} \!+\! \text{h.c.} ,  \notag
\end{align}
where we remind that the induced tunneling amplitude $\mathcal{J}_x^{\text{eff}}$ and the Peierls phase-factors $\phi_{m,n}$ are given in \eqref{induced_hop_peierls}. In particular, since all the plaquettes $\square = \square^{+}$,  the flux is uniform over the lattice $\Phi=\Phi^{+}$, see Eq. \eqref{flux_general}. Using this Wannier-Stark-ladder scheme, uniform-flux configurations were realized in Munich \cite{Aidelsburger:2013} and at the MIT \cite{Ketterle:2013}, with fluxes $\Phi=1/4$ and $\Phi=1/2$, respectively.

Having obtained the effective Hamiltonian $\hat{\mathcal H}_{\rm eff}$, which describes the motion on a square lattice pierced by an effective  uniform flux $\Phi$ per plaquette, we now evaluate the micro-motion for the Wannier-ladder scheme. The micro-motion operator is readily obtained by setting $s(m)=m$ in Eq. \eqref{micro_motion_simple}. Considering that $\omega t \gg \kappa/\hbar \omega $, the main contribution to the micro-motion operator reads
\be
\hat{\mathcal {M}} (t) \approx \omega t  \sum_{m,n} \hat n_{m,n}  \, m \,  \equiv \omega t \hat x /a,
\ee
and we point out that this contribution stems from the superlattice function $s(m)=m$ associated with the Wannier-Stark ladder, see Eq. \eqref{micro_motion_simple}. In this case, the full time-evolution operator  \eqref{full_description} can be approximated as 
\be
\hat U (t; t_0) \approx e^{-i \omega t \hat x/a} \hat U_{\text{eff}} (t; t_0) , \label{full_description_wannier}
\ee
where we remind that $\hat U_{\text{eff}} (t; t_0)$ essentially describes the long-time dynamics due to the effective Hamiltonian, preceded by the initial kick $\exp [i \hat{\mathcal {M}} (t_0)]$. The result \eqref{full_description_wannier} expresses the fact that the micro-motion is associated with a drift $\Delta k_x=\omega t/a$ in quasi-momentum space.  Hence, if the system is initially prepared in the ground-state of the effective Hamiltonian,  the corresponding momentum-density peaks \cite{Gerbier:2010} will travel across the first Brillouin zone (FBZ) within each period of the driving $T$. Note that this micro-motion does not depend on the wave-vector difference ${\bs q}$ in Eq. \eqref{mod_munich_one}, and in this sense, it exists for all values of the effective flux $\Phi$. For $\Phi=1/4$, one can define the magnetic FBZ as $k_{x,y} \in [- \pi/2a , \pi/2a [$. This reduction of the FBZ, together with the fact that $\exp [i \hat{\mathcal {M}} (t)]$ is time-periodic with period $T$, implies that the peaks will travel \emph{twice} across the FBZ  during each period. This rapid motion in momentum space, which can be interpreted as Bloch oscillations due to the strong gradient, potentially complicates any analysis based on time-of-flight-images. Note that this micro-motion is similar to the situation encountered in the well-known  \emph{off-resonant}-shaken 1D optical lattice  \cite{Lignier:2007,Arimondo2012}, where the micro-motion operator reads $\hat{\mathcal {M}} (t)=(\kappa/\hbar \omega a) \hat x \sin (\omega t)$; here $\kappa$ and $\omega$ also denote the modulation amplitude and frequency respectively, see Ref. \cite{goldmandalibard}. However, in the latter case, the position of the momentum-density peaks \emph{oscillate} within a period, instead of performing a constant drift. 

Additionally, we  observe that the micro-motion described by the full operator $\hat{\mathcal {M}} (t) $ in Eq. \eqref{micro_motion_simple}  is also accompanied with discrete kicks $\Delta {\bs k}= \pm \bs q \times \text{integer}$, which generate additional peaks in the momentum distribution. We find that the amplitude of all these momentum peaks oscillate within the micro-motion, in addition to the drift $\Delta k_x=\omega t/a$ described above. The generation of additional peaks due to $\hat{\mathcal {M}} (t) $ will be illustrated in the next Section \ref{section_two_site}. 

The full-time dynamics of the modulated Wannier-Stark ladder has been recently investigated by Bukov and Polkovnikov in Ref. \cite{Bukov:2014}, where a special emphasis has been set on the wiggling cyclotron orbits undergone by atoms at the plaquette level.

\subsection{Two-site square superlattices and local addressing of the tunneling} \label{section_two_site}

We now perform the same analysis for the two-site superlattice, Fig. \ref{Fig_Schema} (a), for which $s(m)=\nicefrac{1}{2} (-1)^m$. In contrast with the Wannier-Stark ladder, this superlattice displays both types of plaquettes $\square^{\pm}$, arranged in alternating columns. Hence, considering the time-modulation in Eq. \eqref{mod_munich_one}, we find that the effective Hamiltonian $\hat{\mathcal{H}}_{\text{eff}}$ is of the form \eqref{ham_stag_1}, where the indices $m^{\pm}$ correspond to alternating columns along the $x$ direction. As discussed in Section \ref{section_main_mod}, the hopping amplitude $\mathcal{J}_x^{\text{eff}}$, the Peierls phase-factors $\phi_{m,n}$ and the fluxes per plaquette $\Phi^{\pm}$ are given by Eqs. \eqref{induced_hop_peierls}-\eqref{flux_general}. Altogether, in the present case, the flux pattern is staggered, with flux $2 \pi \Phi^{\pm}=\pm q_y /a$ in alternating columns, see Eqs. \eqref{flux_plus}-\eqref{flux_minus} and Refs. \cite{Aidelsburger:2011,Aidelsburger:2013b}. 

This staggered-flux configuration, which naturally arises in the two-site superlattice, can be rectified so as to generate a uniform-flux pattern over the whole superlattice. This can be realized by modifying the simple  modulation's spatial dependence $v(m,n)$ in Eq. \eqref{mod_munich_one_bis}. Specifically, we consider the  following two-fold partition,
\begin{align}
&v(m,n)=\frac{1}{2} \Biggl \{ f_1 (m) e^{i g_1 (n)} + f_2 (m) e^{i g_2 (n)}      \Biggr \}, \label{split_mod}\\
&f_1(m+1)-f_1(m) \ne 0, \, \, \, \, f_2(m+1)-f_2(m) = 0 \, \text{ for $m^{+}$} , \notag \\
&\, f_1(m+1)-f_1(m) = 0, \, \, \, \, f_2(m+1)-f_2(m) \ne 0 \, \text{ for $m^{-}$} . \notag
\end{align}
In this case, the restored tunneling is independently addressed in alternating columns, see Eq. \eqref{general_restored}. In particular, the flux associated with alternating columns $m^{\pm}$ are individually controlled by the functions $g_{1,2} (n)$, respectively.  Let us consider the specific functions
\begin{align}
&f_1(m)= \cos \left (m \frac{\pi}{2} - \frac{\pi}{4} \right) , \, g_1 (n) = - n \pi/2 , \notag \\
& f_2(m)= \cos \left (m \frac{\pi}{2} + \frac{\pi}{4} \right), \, g_2 (n) = (n-1) \pi/2 , \label{split_munich}
\end{align}
which correspond to the scheme implemented in Ref. \cite{Aidelsburger:2014}.  These satisfy the conditions in Eq. \eqref{split_mod} and generate a uniform flux $2 \pi \Phi=+\pi/2$ over the whole lattice. Note that the two functions $g_{1,2}$ are typically associated with distinct pairs of laser beams \cite{Aidelsburger:2014}.

It is straightforward to generalize this local-addressing procedure to more complex superlattices, in which case the function $v(m,n)$ in Eq. \eqref{split_mod} could be split into more different parts. Additionally, we point out that this local addressing could also be exploited to generate other flux patterns, potentially richer than the staggered or uniform patterns.\\

\begin{figure}[h!]
\includegraphics[width=9.5cm]{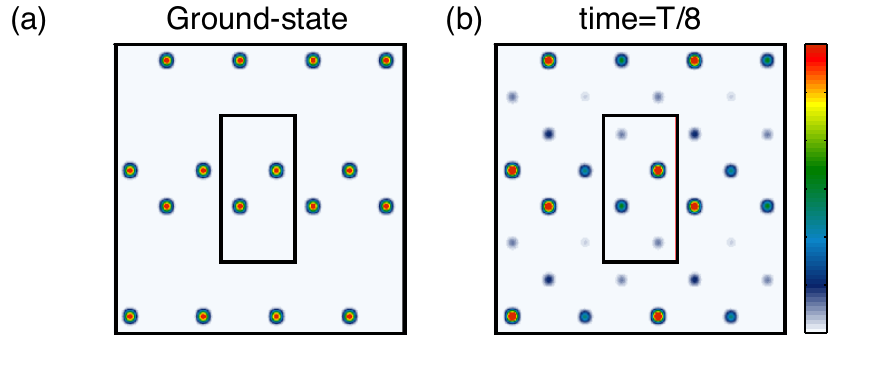}
\vspace{-0.cm} \caption{(a) Momentum distribution associated with the ground-state $\vert \text{GS} \rangle $ of the effective Hamiltonian $\hat{\mathcal{H}}_{\text{eff}}$, corresponding to a staggered-flux lattice with flux $\Phi^{\pm}=\pm 1/4$. (b) Momentum distribution of the time-evolved state  $\vert \psi (t) \rangle$, at time $t=T/8$, when initially starting the evolution with the ground-state $\vert \psi (t_0) \rangle =\vert \text{GS} \rangle $. The amplitude of the density peaks oscillate within a period of the driving, but the position of the peaks remain constant. Here $\kappa=8J$ and $\omega=20J$, where $J$ is the hopping amplitude.  This latter figure is to be compared with the experimental data shown in Fig. 2b in Ref. \cite{Aidelsburger:2011}, which corresponds to $\Phi^{\pm}=\pm 1/4$ and $2\kappa/(\hbar\omega) \approx 0.48$. The reduced FBZ, $k_{x} \in [- \pi/2a , \pi/2a [$ and $k_{y} \in [- \pi/a , \pi/a [$, is highlighted in all figures. }\label{Fig_Micro}
\end{figure}

We now discuss the micro-motion in the driving schemes involving a two-site superlattice. First, let us consider the simple time-modulation in Eq. \eqref{mod_munich_one} leading to the staggered-flux configuration. Using Eq. \eqref{micro_motion_simple}, we obtain the micro-motion operator 
\be
\hat{\mathcal {M}} (t) = \sum_{m,n} \hat n_{m,n} \left \{   \frac{\omega t}{2} (-1)^{m} + \frac{2 \kappa}{ \hbar \omega} \sin \left [ \omega t + \bs{q}\cdot \bs{R}   \right ]     \right \}.\label{micro_two_site}
\ee
The first part in Eq. \eqref{micro_two_site}, which is due to the superlattice $s(m)=\nicefrac{1}{2} (-1)^{m}$, is essentially trivial. Thus, focusing on the second part in Eq. \eqref{micro_two_site}, we find that the effects attributed to the micro-motion operator mainly consists in kicks $\Delta {\bs k}= \pm \bs q \times \text{integer}$, potentially generating new peaks in quasi-momentum space.  This is similar to the situation encountered in the Wannier-Stark ladder in Section \ref{section_wannier_ladder}. However, in contrast to the latter case, the momentum peaks do not travel in $\bs{k}$-space within a period: these peaks are well defined at specific $\bs k$ points, \emph{at all times}, and they simply oscillate in amplitude over each period, as described in Section \ref{section_kick_two_site}. We illustrate this effect in Fig. \eqref{Fig_Micro},  where we show the time-evolved momentum distribution, starting with the ground-state $=\vert \text{GS} \rangle$ of the effective Hamiltonian $\hat{\mathcal{H}}_{\text{eff}}$. Here the effective staggered-flux is chosen to be $2 \pi \Phi^{\pm}=\pm\pi/2$, as in the experiment \cite{Aidelsburger:2011}, namely $q_x=q_y=\pi/2a$. The corresponding ground-state's momentum distribution displays two sharp peaks within the FBZ, see Fig. \ref{Fig_Micro}(a).  We set the initial state to be this ground-state, $\vert \psi (t_0) \rangle =\vert \text{GS} \rangle $, and compute the momentum distribution of the evolving state $\vert \psi (t) \rangle $ according to the time-evolution operator in Eq. \eqref{full_description}. Figure \ref{Fig_Micro}(b) shows the corresponding distribution at some arbitrary time $t=T/8$, where two additional peaks are observed. As discussed above, these two extra peaks are obtained by translating the two initial peaks according to the vectors $\Delta {\bs k}= \pm \bs{q} = \pm (\pi/2a) (\bs{1}_x + \bs{1}_y)$.  A comparison with the time-of-flight image shown in Ref. \cite{Aidelsburger:2011} indicates that this four-peaks pattern is indeed robust and observable in experiments. In summary, we point out that the experimental data cannot be interpreted in terms of the effective Hamiltonian $\hat{\mathcal{H}}_{\text{eff}}$ alone, as some of its features are specifically captured by the micro-motion operator $\hat{\mathcal {M}} (t)$ in Eq. \eqref{micro_two_site}. 

 To conclude this subsection, we present the expression for the micro-motion operator \eqref{micro_operator} in the case of the rectified-flux scheme, which is based on the time-modulation \eqref{split_mod}-\eqref{split_munich},
\begin{widetext}
\begin{align}
\hat{\mathcal {M}} (t) &=\sum_{m,n} \hat n_{m,n} \left [ \frac{\omega t}{2} (-1)^{m} + \frac{2 \kappa}{ \hbar \omega}  \Biggl \{ \sin \left [ \omega t + g_1 (n)    \right ] f_1 (m) + \sin \left [ \omega t + g_2 (n) \right ] f_2 (m)   \Biggl \}     \right ] \notag \\
&=\sum_{m,n} \hat n_{m,n} \left [ \frac{\omega t}{2} (-1)^{m} + \frac{2 \kappa}{ \hbar \omega}  \Biggl \{ \sin \left [ \omega t - \frac{\pi}{2} n    \right ] \cos \left [ \frac{\pi}{2} m -\frac{\pi}{4}   \right ] + \sin \left [ \omega t + \frac{\pi}{2} (n    -1) \right ] \cos \left [ \frac{\pi}{2} m + \frac{\pi}{4}   \right ]   \Biggl \}     \right ].
\end{align}
This micro-motion operator leads to a similar behavior as the one related to the staggered-flux scheme [Eq. \eqref{micro_two_site} and Fig. \ref{Fig_Micro}]: the momentum-distribution of the evolving ground-state shows a series of peaks shifted by $\Delta k =\pi/2a$ along both directions; these peaks oscillate in amplitude, but their position in $k$-space remains constant at all times. Again, this is in sharp contrast with the Wannier-Stark-ladder case, where the distribution peaks travel in $k$-space within a period of the driving [Section \ref{section_wannier_ladder}]. Consequently, we have demonstrated that two schemes leading to a uniform-flux configuration -- the uniformly modulated Wannier-Stark ladder and the locally modulated two-site superlattice -- could present radically different micro-motions. 
\end{widetext}

\subsection{Modulated honeycomb lattices: a momentum-space analysis}\label{section_honey}

We finally present an explicit application of the momentum-space approach presented in Section \ref{section_nigel}, by analyzing the dynamic superlattice introduced in Ref.~\cite{Baur:2014ux}. This modulated optical-lattice scheme generates a model closely akin to the Haldane model \cite{Haldane:1988}, a lattice model displaying nearest-neighbor (NN) and next-nearest-neighbor (NNN) hopping terms, and leading to non-trivial Chern bands. 

This scheme involves a static optical lattice formed from three in-plane waves, with wave vectors 
\be
 {\bm \kappa}_1 = \kappa(0,1), \, {\bm \kappa}_2 = \kappa(-\sqrt{3}/2,-1/2) , \, {\bm \kappa}_3 = \kappa(\sqrt{3}/2,-1/2),\notag 
\ee
which is designed to form a distorted honeycomb lattice, where the two inequivalent sites A and B are separated by an energy offset $\hbar\delta = \hbar\omega + \hbar\delta'$, see Ref. \cite{Baur:2014ux}; note that we allow detuning from resonance, $\delta'\neq 0$. This configuration determines two sectors, $\alpha = \{ 0 , 1 \}$, which are associated with the sites of type A and B, respectively; see Section \ref{section_nigel}. The reciprocal lattice vectors are the momentum transfers ${\bm K}_{ij} = {\bm \kappa}_i - {\bm \kappa}_j$, from which one can construct the next-nearest-neighbor (i.e. A-A and B-B) vectors, 
\be
{\bm a}_1 = \frac{2\pi}{3\kappa}(\sqrt{3},-1), \, {\bm a}_2 = \frac{4\pi}{3\kappa}(0,1), \, {\bm a}_3 = \frac{2\pi}{3\kappa}(-\sqrt{3},1).
\ee 
The static Hamiltonian is characterized by the energy offset, $\hbar\delta$, the nearest-neighbor (A-B) tunneling, $t_{01}$, and the next-nearest neighbor tunnelings, $t_{00}$ (A-A) and $t_{11}$ (B-B). Working to second order in  $t_{01}$ [Eqn. (\ref{eq:2site1})], the static Hamiltonian leads to dispersions for particles moving on the (decoupled) A and B sublattices of
\begin{eqnarray}
\label{eq:e0_honey}
\epsilon_0({\bm k}) & = & -3 \frac{t_{01}^2}{\hbar\omega} - \left(t_{00} + \frac{t_{01}^2}{\hbar\omega}\right)  f_{\Delta}({\bm k}), \\
\epsilon_1({\bm k}) & = & \hbar\delta' +3 \frac{t_{01}^2}{\hbar\omega} - \left(t_{11} - \frac{t_{01}^2}{\hbar\omega}\right)  f_{\Delta}({\bm k}) ,
\label{eq:e1_honey}
\end{eqnarray}
where $f_\Delta({\bm k}) \equiv 2\sum_{i=1}^3 \cos({\bm a}_i\cdot {\bm k})$ is the characteristic dispersion for a triangular lattice. 
The A-B tunneling is restored by a dynamic modulation of the potential on the B sites\cite{Baur:2014ux}
 \begin{equation}
\hat{H}_{11}^{(1)} = V_{\rm D}  \left (  e^{i{\bm \kappa}_1\cdot {\bm r}} + \text{j} e^{i{\bm \kappa}_2\cdot {\bm r}} + \text{j}^2 e^{i{\bm \kappa}_3\cdot {\bm r}} \right ) \hat{P}^1 , \quad \text{j}=e^{i \frac{2\pi}{3}} .\notag
\end{equation}
This causes wave vector transfers of ${\bm q}={\bm \kappa}_{j={1,2,3}}$, all of which are equivalent since ${\bm \kappa}_i-{\bm \kappa}_j$ are reciprocal lattice vectors, see Section \ref{section_nigel}. An explicit calculation of the terms (\ref{eq:2site2}) leads to 
\begin{align}
\label{eq:v01_honey}
v_{10}({\bm k})  \equiv   \langle 1, {\bm k}+ {\bm q}| \frac{-1}{\hbar\omega}\hat{H}_{11}^{(-1)}\hat{H}_{10}^{(0)}|0, {\bm k}\rangle = \frac{t_{10} V_{\rm D}}{\hbar\omega} f_{\rm hc}({\bm k}),
\end{align}
where $f_{\rm hc}({\bm k}) \equiv \sum_i \exp(-i {\bm k}\cdot {\bm R}_i)$ is the characteristic dispersion for the honeycomb lattice with nearest neighbour (A-B) vectors 
\begin{align}
&{\bm R}_{1} = \frac{4\pi}{3\sqrt{3}\kappa}(1,0), \quad  {\bm R}_{2} = \frac{4\pi}{3\sqrt{3}\kappa}(-1/2,\sqrt{3}/2), \notag \\
& {\bm R}_{3} = \frac{4\pi}{3\sqrt{3}\kappa}(-1/2,-\sqrt{3}/2).\notag
 \end{align}
Combining the dispersions \eqref{eq:e0_honey}-\eqref{eq:e1_honey} with the coupling (\ref{eq:v01_honey}) in the effective Hamiltonian \eqref{eq:2site_heff} gives a complete description of the band structure for this model, and allows the topology of the bands to be readily determined. Indeed, for a $2 \times 2$ Hamiltonian matrix of the form \eqref{eq:2site_heff}, the Chern number of the bands \cite{Thouless1982,Kohmoto:1989} can be simply evaluated by analyzing the vortex structure associated with singular points ${\rm K}_{\pm}$ of the Brillouin zone, where the (complex) off-diagonal component $v_{10}$ vanishes \cite{Haldane:1988,Goldman:2013}. Here, we find that the complex function $v_{10}$ in Eq. \eqref{eq:v01_honey} indeed vanishes at the special points ${\rm K}^+$ and ${\rm K}^-$, located at ${\bm k}_{\rm K^+} = k(0,1)$ and ${\bm k}_{\rm K^-} = k(0,-1)$, respectively, and that it accumulates a phase $\pm 2\pi$ when circulating around them.  If the difference of the diagonal elements $\epsilon_0({\bm k}) - \epsilon_1({\bm k}+{\bm q})$ is non-zero and of opposite sign at these points  ${\bm k}_{\rm K}$ and ${\bm k}_{\rm K'}$, then a spectral gap opens and the two separated energy bands will have Chern numbers of $+1$ and $-1$, see Refs. \cite{Haldane:1988,Goldman:2013}. It is straightforward to show that  the physical parameters can be chosen  to achieve this goal.

\section{Generalization to schemes using more driving frequencies}\label{section_more}

In the previous Section \ref{section_two_site}, we have shown how the restoration of the tunneling can be controlled locally by tailoring the spatial function $v(m,n)$, which characterizes the time-periodic modulation $\hat V (t)$, Eq. \eqref{mod_def}. Another strategy consists in designing static superlattices with higher-order energy offsets, $\Delta_{N}= N \Delta$ where $N>1$ are some integers, as illustrated in Fig. \ref{Fig_Schema} (b) for $N=2$, see also Ref. \cite{Kennedy:2013}. In this scenario, links associated with an offset $\pm \Delta_{N}$ are re-activated by higher-harmonic components of the time-periodic modulation $\hat V (t)$, with resonant frequency $\omega_{N} = N \omega$; here, we keep $\omega=\Delta/\hbar$ as the fundamental harmonics, so that $\hat V (t+T)=\hat V (t)$ with $T=2 \pi/\omega$. Importantly,  to first order in the amplitude of the dynamic modulation, links associated with different offsets can be addressed \emph{individually}, which is due to the fact that the effective Hamiltonian $\hat{\mathcal{H}}_{\text{eff}}$ in Eq. \eqref{effective_ham_two} has decoupled contributions from the different harmonics, at this order of the perturbative expansion \cite{goldmandalibard}. 

\subsection{Higher-order energy offsets and driving frequencies}

We now explicitly show how the formalism of Section \ref{section_one_harmonics} generalizes to this situation. First, we extend our site-labeling notations [Eqs. \eqref{def_m}-\eqref{def_m_2}] as
\be
\delta_s (m^{+j}) = +j , \qquad \delta_s (m^{-j}) = -j, \label{def_m_j}
\ee
where  $\delta_s (m) = s(m+1) - s(m) $,  $j>0$ is an arbitrary integer, and where we remind that $s(m)$ denotes the superlattice function [Eq. \eqref{statone}]. Namely, the neighboring sites $(m^{+j},n)$ and $(m^{+j}+1,n)$ are now allowed to be separated by an energy offset $\Delta_j = + j \Delta$. We then include higher harmonics in the time-modulation 
\be
\hat V (t) = \sum_{j\ne 0} \hat H^{(+j)} \exp (i j \omega t),
\ee 
where the components are defined as
\begin{align}
&\hat H^{(+j)}= \kappa \sum_{m,n} \hat n_{m,n} v_j (m,n) = \left [ \hat H^{(-j)}  \right]^{\dagger} , \quad \omega = \Delta / \hbar .  \label{mod_def_bis}
\end{align}
Performing the unitary transformation \eqref{transf_hof} and using Eq. \eqref{ch_frame_2}, we obtain the transformed time-dependent Hamiltonian
\begin{align}
&\hat {\mathcal H} (t) = \hat {\mathcal H}^{(0)} + \sum_{j>0} \hat {\mathcal H}^{(+j)} e^{i j\omega t} + \hat {\mathcal H}^{(-j)} e^{-i j\omega t} ,\label{ham_dec_bis}\\
& \hat {\mathcal H}^{(0)} =\hat T_y + \hat U_{\text{onsite}}, \notag \\
&\hat {\mathcal H}^{(+j)} = \kappa \sum_{m,n} \hat n_{m,n}\,  v_j (m,n) \notag \\
&\qquad \, - J_x \left \{ \sum_{m^{+j},n} \hat a_{m+1,n}^{\dagger} \hat a_{m,n} + \sum_{m^{-j},n} \hat a_{m,n}^{\dagger} \hat a_{m+1,n}\right \} = [\hat {\mathcal H}^{(-j)}]^{\dagger}, \notag
\end{align}
which straightforwardly generalizes the Hamiltonian in Eq. \eqref{ham_dec}. 

The effective Hamiltonian $\hat{\mathcal H}_{\rm eff}$ is computed using the general definition in Eq. \eqref{effective_ham_two}, which yields
\begin{align}
\hat{\mathcal H}_{\rm eff}= \hat T_y + \hat U_{\text{onsite}}- \frac{J_x \kappa}{\hbar \omega}& \sum_{j>0} \frac{1}{j}
\Biggl \{ \sum_{m^{+j},n} \hat a_{m+1,n}^{\dagger} \hat a_{m,n} \gamma_j (m,n)  \notag \\
&- \sum_{m^{-j},n} \hat a_{m+1,n}^{\dagger} \hat a_{m,n} \gamma^*_j (m,n) + \text{h.c.} \Biggr \},\notag
\end{align}
where $\gamma_j (m,n) = v_j^*(m,n) - v_j^*(m+1,n)$. This shows that links associated with different offsets $\Delta_j=j \Delta$ are individually addressed by the related harmonic components. Assuming that $\gamma_j (m,n) = \rho_j \, e^{i \phi_{m,n}^j}$, we obtain a simple form for the effective Hamiltonian
\begin{align}
\hat{\mathcal H}_{\rm eff}= \hat T_y + \hat U_{\text{onsite}} - \sum_j \mathcal{J}_{x,j}^{\text{eff}} &\Biggl \{ \sum_{m^{+j},n}   \hat a_{m+1,n}^{\dagger} \hat a_{m,n} e^{i \phi_{m,n}^j}\notag \\
& \!-\! \sum_{m^{-j},n}   \hat a_{m+1,n}^{\dagger} \hat a_{m,n} e^{-i  \phi_{m,n}^j } \!+\! \text{h.c.} \Biggr \} ,  \label{general_hopping}
\end{align}
where the induced tunneling amplitude is now given by 
\be
\mathcal{J}_{x,j}^{\text{eff}}= J_x \kappa \rho_j / j\hbar \omega .\label{eq:inhomog}
\ee

Let us apply this scheme to the three-site superlattice \cite{Kennedy:2013} shown in Fig. \ref{Fig_Schema} (b). In this case, the sites are all of type $m=m^{+1}$ or $m=m^{-2}$. Therefore, using Eq. \eqref{general_hopping}, we find that a uniform flux $\Phi$ is readily obtained by considering a two-harmonic modulation satisfying
\be
\phi_{m,n}^1 = -\phi_{m,n}^2, \quad 2 \pi \Phi=\phi_{m,n}^1 - \phi_{m , n+1}^1,
\ee
namely, a moving potential of the form
\begin{align}
\hat V (t)&\!=\! 2 \kappa \sum_{m,n} \hat n_{m,n} \left \{ \cos \left ( \omega t + \bs{q} \cdot \bs{R}  \right ) + 2 \cos \left (2 \omega t - \bs{q} \cdot \bs{R}  \right )  \right \}. \notag  
\end{align}
Note the additional factor of two in the second term, which allows to restore a uniform tunneling amplitude $\mathcal{J}_{x}^{\text{eff}}$ over the lattice, see Eq. \eqref{eq:inhomog}. \\

\subsection{Driving the hopping along $x$ and $y$}

Finally, we discuss the possibility to induce and control the tunneling matrix elements along \emph{both} spatial directions. This can be realized by adding two superlattices, one for each direction. In order to address these two directions individually, we follow the same reasoning as above and consider different energy offsets\begin{align}
&\hat S= \Delta \sum_{m,n}   \hat n_{m,n} \left \{ s_x(m) + s_y(n) \right \} , \\
&s_x(m+1)-s_x(m)= \pm 1 , \quad s_y(n+1)-s_y(n)= \pm 2.
\end{align}
Performing the unitary transformation \eqref{transf_hof}, we obtain the transformed Hamiltonian
\begin{align}
\hat {\mathcal H} (t) = \hat{\mathcal{T}}_x + \hat{\mathcal{T}}_y + \hat U_{\text{onsite}} + \hat V (t), \label{ch_frame_bis}
\end{align}
where the hopping terms are now modified along both directions:
\begin{align}
&\hat{\mathcal T}_x = -J_x \sum_{m,n} \hat a_{m+1,n}^{\dagger} \hat a_{m,n} e^{i \omega t \left [ s_x(m+1) - s_x(m)   \right ]} + \text{h.c.} \\
&\hat{\mathcal T}_y = -J_y \sum_{m,n} \hat a_{m,n+1}^{\dagger} \hat a_{m,n} e^{i \omega t \left [ s_y(n+1) - s_y(n)   \right ]} + \text{h.c.}
\end{align}
In order to restore the hopping, we consider a two-harmonic modulation of the form 
\begin{align}
\hat V (t)&\!=\! \kappa \sum_{m,n} \hat n_{m,n} \left \{ v_1 (m,n) e^{i \omega t} + v_2 (m,n) e^{i 2 \omega t}+ \text{h.c.}     \right \}  \label{mod_def_tri} .
\end{align}
\begin{widetext}
The time-dependent Hamiltonian is then written as
\begin{align}
&\hat  {\mathcal H} (t) = \hat  {\mathcal H}^{(0)} +\hat  {\mathcal H}^{(+1)} e^{i \omega t} + \hat  {\mathcal H}^{(-1)} e^{-i \omega t} + \hat  {\mathcal H}^{(+2)} e^{i 2 \omega t} + \hat  {\mathcal H}^{(-2)} e^{-i 2 \omega t} ,\notag\\
& \hat  {\mathcal H}^{(0)} =  \hat U_{\text{onsite}}, \notag \\
&\hat  {\mathcal H}^{(+1)} = \kappa \sum_{m,n} \hat n_{m,n}\,  v_1 (m,n)  - J_x \left \{ \sum_{m^+,n} \hat a_{m+1,n}^{\dagger} \hat a_{m,n} + \sum_{m^-,n} \hat a_{m,n}^{\dagger} \hat a_{m+1,n}\right \} = [\hat  {\mathcal H}^{(-1)}]^{\dagger}. \notag \\
&\hat  {\mathcal H}^{(+2)} = \kappa \sum_{m,n} \hat n_{m,n}\,  v_2 (m,n)  - J_y \left \{ \sum_{m,n^+} \hat a_{m,n+1}^{\dagger} \hat a_{m,n} + \sum_{m,n^-} \hat a_{m,n}^{\dagger} \hat a_{m,n+1}\right \} = [\hat  {\mathcal H}^{(-2)}]^{\dagger}, \notag
\end{align}
where we used the site-labeling convention $\delta_{s_x}(m^{\pm})=\pm1$ and $\delta_{s_y}(n^{\pm})=\pm 2$, which generalizes Eqs. \eqref{def_m}-\eqref{def_m_2} to two spatial directions.

Finally, the effective Hamiltonian $\hat{\mathcal H}_{\rm eff}$ is computed using the general definition in Eq. \eqref{effective_ham_two}, and it reads

\begin{align}
\hat{\mathcal H}_{\rm eff}=\hat U_{\text{onsite}}&- \frac{J_x \kappa}{\hbar \omega}
\Biggl \{ \sum_{m^{+},n} \hat a_{m+1,n}^{\dagger} \hat a_{m,n} \gamma_1 (m,n) - \sum_{m^{-},n} \hat a_{m+1,n}^{\dagger} \hat a_{m,n} \gamma^*_1 (m,n) + \text{h.c.} \Biggr \} \notag \\
&- \frac{J_y \kappa}{2 \hbar \omega}
\Biggl \{ \sum_{m,n^+} \hat a_{m,n+1}^{\dagger} \hat a_{m,n} \gamma_2 (m,n) - \sum_{m,n^{-}} \hat a_{m,n+1}^{\dagger} \hat a_{m,n} \gamma^*_2 (m,n) + \text{h.c.} \Biggr \},\label{effective_both_direction}
\end{align}
where  
\be
\gamma_{1} (m,n) = v_{1}^*(m,n) - v_{1}^*(m+1,n), \quad  \gamma_{2} (m,n) = v_{2}^*(m,n) - v_{2}^*(m,n+1).\label{def_two_harm_diff}
\ee
The result in Eqs. \eqref{effective_both_direction}-\eqref{def_two_harm_diff} shows how the tunneling matrix elements associated with the two spatial directions can be individually controlled by the two different harmonic components of the time-modulation in Eq. \eqref{mod_def_tri}. This can potentially generate very rich flux patterns in two dimensional lattice systems.  Generalization to three dimensions is straightforward, as it would simply require a superlattice along $z$, together with an additional (resonant) harmonic component in the time-modulation $\hat V (t)$. 
\end{widetext}

\section{Conclusions} \label{section_conclusions}

This work proposed a framework to investigate the physics of time-periodic modulated systems presenting resonant features. Rooted in the formalism of Refs. \cite{goldmandalibard,Rahav:2003}, this novel approach offers a systematic way to calculate the effective Hamiltonian $\hat H_{\text{eff}}$ and kick operator $\hat K (t)$  for such resonant-driving situations. The motivations for obtaining the effective Hamiltonian and its corresponding (Floquet) spectrum is well established \cite{Kitagawa:2010,Arimondo2012,Cayssol:2013}, however, the effects associated with the kick operator are also found to be crucial for the analysis of driven systems \cite{goldmandalibard,Bukov:2014_review,Bukov:2014}. In particular, the micro-motion can potentially produce large and rapid oscillations of experimental observables -- e.g. momentum distributions or spin populations -- precluding any instructive measurement of these quantities. In this work, we highlighted the simple but important fact that the micro-motion can be \emph{different} for time-modulated systems leading to the \emph{same} effective Hamiltonian. This aspect was illustrated by comparing the modulated Wannier-Stark ladder [Section \ref{section_wannier_ladder}] with the modulated two-site square superlattice [Section \ref{section_two_site}], for which the time-evolving peaks in the momentum distribution showed drastically different behaviors. 

This work also showed the possibility to generate a large variety of flux patterns in two-dimensional lattice systems through the local restoration and control of tunneling. This can be achieved by tailoring the space-dependent features of the time-modulation [Section \ref{section_two_site}] and/or using superlattices with different energy offsets [Section \ref{section_more}]. Applying these schemes to more spatial directions and spin structures offers a versatile toolbox to generate a wide variety of lattice models and gauge fields, suggesting interesting avenues in the field of quantum simulation.

We stress that a single-band tight-binding approximation has been assumed in the examples considered in this work. We note that multiband systems subjected to periodic modulations could permit multiphoton processes that promote atoms to high-energy untrapped states, endangering the stability of these engineered models at very long times.

The interplay between time-periodic modulations and inter-particle interactions is conjectured to be the source of  heating in experiments \cite{Ketterle:2013,Jotzu:2014,Aidelsburger:2014}. Recently, several works investigated the effects of interactions in time-modulated lattices \cite{Abdullaev:2003,Bilitewski:2014tt,Creffield:2008cp,Choudhury:2014ch,Choudhury:2014tu}, where regimes of dynamical instabilities were identified. The thermodynamics of driven systems was also explored in Refs. \cite{DAlessio:2013fv,Langemeyer:2014ww,Lazarides:2013uh,DAlessio:2014,Lazarides:2014,Ponte:2015,Cuetara:2015}.  A general understanding of these heating sources still constitutes an important issue to be addressed in this framework, for instance, in view of creating novel (topological) strongly-correlated states with cold-atoms. \\

 During the completion of the revised manuscript, we became aware of a similar work by Eckardt and Anisimovas \cite{Anisimovas_2015}, where equivalent expressions for the effective Hamiltonian and micro-motion operators were obtained through an alternative method.

\begin{acknowledgments}
The authors are pleased to acknowledge  I. Bloch, M. Bukov, S. Nascimb\`ene, U. Schneider and D.-T. Tran for stimulating inputs. N.G. is supported by the Universit\'e Libre de Bruxelles (ULB) and the FRS-FNRS (Belgium). This research was also funded by IFRAF and ANR (AGAFON), NIM, the EU (SIQS), the Royal Society of London,  and by the European Research Council Synergy Grant UQUAM. M. A. was additionally supported by the Deutsche Telekom Stiftung, and N.C. by EPSRC Grant EP/K030094/1. \\
\end{acknowledgments}

\appendix

\section*{Appendix: Second-order corrections}\label{appendix_second}

In this appendix, we provide second-order corrections to the effective Hamiltonian $\hat{\mathcal H}_{\rm eff}$ obtained in Section \ref{section_effective}. Following Ref. \cite{goldmandalibard}, the second-order corrections to the general effective Hamiltonian in Eq. \eqref{effective_ham_two} are given by 
\begin{align}
\mathcal{C}^{(2)} = \frac{1}{2 (\hbar \omega)^2 } \sum_{j>0} \frac{1}{j^2} \bigl [ \bigl [\hat {\mathcal H}^{(+j)} , \hat {\mathcal H}^{(0)} \bigr ] , \hat {\mathcal H}^{(-j)} \bigr ]    \!+\! \text{h.c.} ,\label{effective_ham_second_order} 
\end{align} 
where the operators $\hat {\mathcal H}^{(j)}$ were defined in Section \ref{section_general}. 

We now apply this expression \eqref{effective_ham_second_order} to the specific operators in Eq. \eqref{ham_dec}. In order to highlight the main effects, we make two simplifications. First we omit the onsite terms $\hat U_{\text{onsite}}$, whose contributions to second-order effects are typically weak; this yields $\hat {\mathcal H}^{(0)} \approx \hat T_y$. Then, since $\kappa \gg J_x$, we approximate 
\be
\hat {\mathcal H}^{(+1)} \approx \kappa \sum_{m,n} \hat n_{m,n}\,  v (m,n).
\ee 
We then find that the main corrections to the effective Hamiltonian $\hat{\mathcal H}_{\rm eff}$ are given by [Eq. \eqref{effective_ham_second_order}]
\be
\mathcal{C}^{(2)} = \frac{\kappa^2 J_y}{(\hbar \omega)^2} \sum_{m,n} \hat a_{m,n+1}^{\dagger} \hat a_{m,n} \vert v(m,n) - v(m,n+1) \vert ^2 + \text{h.c.}, \notag
\ee
which corresponds to a renormalization of the hopping along the $y$ direction. Including the zero-th order term, the total hopping operator along the $y$ direction is modified as
\begin{align}
&\hat T_y \rightarrow \hat{T}^{\rm eff}_y = -J_y \sum_{m,n} \mu (m,n) \hat a_{m,n+1}^{\dagger} \hat a_{m,n} + \text{h.c.}, \notag \\
&\mu (m,n)= 1 -  \left (\frac{\kappa}{\hbar \omega} \right)^2  \, \vert v(m,n) - v(m,n+1) \vert ^2 ,\label{second_tunnel}
\end{align}
where $\mu (m,n)$ captures the possible inhomogeneity of the hopping. As realized in Ref. \cite{Creffield:2013gp}, this may be particularly problematic in schemes where $v(m,n) - v(m,n+1)$ is proportional to one of the spatial coordinates $(m,n)$, in which case the hopping can be strongly reduced in large regions of the system. However, we note that this inhomogeneity effect is limited for the schemes discussed in Sections \ref{section_wannier_ladder} and \ref{section_two_site}.  Finally, we note that the corrections in Eq.~\eqref{second_tunnel} are in agreement with the strong-driving result in Eq.~\eqref{amplitudes_strong}.


\begin{thebibliography}{81}
\bibitem[]{footnote} The diverging term $\hat S$ in the static Hamiltonian \eqref{statone} is indeed of the form introduced in Section \ref{section_general}, see Eq. \eqref{eq:hoffset}. We start by writing $\hat S = \hbar \omega \sum_{\alpha} \alpha \hat N_{\alpha}$ ,  where $\hat N_{\alpha}=\sum_{j \in \alpha} \hat n_j $ is the number operator for the sector $\alpha$, and $\hat n_j$ is the number operator at the lattice site $j=(m,n)$; all the sites $j \in \alpha$ belonging to a same sector are determined by the superlattice function $s(m)$. Expanding the number operator $\hat N_{\alpha}$ in terms of projectors $\hat N_{\alpha} = \sum_{n_{\alpha}} n_{\alpha} \hat{P}_{n_{\alpha}} $  where $\hat N_{\alpha} \vert n_{\alpha} \rangle = n_{\alpha}  \vert n_{\alpha} \rangle $, we obtain $\hat S = \hbar \omega \sum_{\alpha , n_{\alpha}}  ( \alpha n_{\alpha} ) \hat{P}_{n_{\alpha}}$.
\end{thebibliography}
\bibliographystyle{apsrev}

 \end{document}